\documentclass[12pt, a4paper]{article}
\usepackage{graphicx}
\usepackage{amssymb}
\usepackage{amsmath}
\usepackage{bm}
\usepackage{color}
\usepackage{theorem}
\usepackage{subcaption}
\usepackage{listings}
\usepackage{colortbl}
\usepackage{tabularx}
\usepackage{longtable}
\usepackage[utf8]{inputenc}
\usepackage[T1]{fontenc}
\usepackage{lmodern}
\usepackage[top=30truemm,bottom=30truemm,left=25truemm,right=25truemm]{geometry}
\usepackage[sort&compress,numbers, merge]{natbib}
\usepackage{multirow}
\usepackage{here}
\usepackage{braket}
\usepackage[normalem]{ulem}

\definecolor{Orange}{cmyk}{0,0.61,0.87,0}
\definecolor{JungleGreen}{cmyk}{0.99,0,0.52,0}
\definecolor{OliveGreen}{cmyk}{0.64,0,0.95,0.40}
\definecolor{Brown}{cmyk}{0,0.81,1,0.60}
\definecolor{RoyalBlue}{cmyk}{0.71,0.53,0,0.12}
\definecolor{Gray}{cmyk}{0,0,0,0.40}
\definecolor{LightPink}{cmyk}{0.0,0.25,0,0}
\definecolor{LLightPink}{cmyk}{0.0,0.10,0,0}
\definecolor{LightBlue}{cmyk}{0.25,0,0,0}
\definecolor{LightGray}{cmyk}{0,0,0,0.2}

\def\vnr{\Upsilon{}}
\def\mvdot{{\scriptscriptstyle\circ}}
\def\scA{\mathcal A_\perp{}}
\def\tr{{\rm Tr}}
\def\diff{{\rm d}}
\def\e{{\rm e}}
\def\mJ{\mathcal J}

\newcommand{\del}{\partial}

\newcommand{\nn}{\nonumber}

\usepackage[colorlinks=true,linkcolor=OliveGreen,citecolor=RoyalBlue,urlcolor=RoyalBlue]{hyperref}

\renewcommand{\thefootnote}{\fnsymbol{footnote}}
\allowdisplaybreaks[1]

\begin{document}

\begin{titlepage}
\begin{flushright}
{\tt 
UT-HET-143
}
\end{flushright}
\begin{center}

\vspace{1.5cm}

\textbf{\Large
Resummed multi-line gamma-ray spectra for Cherenkov Telescopes from heavy spin-1 dark matter
}
\vspace{0.5cm}

Motoko Fujiwara$^{a,b}$\footnote{
E-mail address: 
\href{mailto:motoko@sci.u-toyama.ac.jp}{\tt motoko@sci.u-toyama.ac.jp}},
Martin Vollmann$^{c}$\footnote{
E-mail address: 
\href{mailto:info@fullman.de}{\tt  Martin@FullMan.de}}

\vskip 0.8cm

\vspace{0.25cm}
{\small  \it 
$^a$Physik-Department, Technische Universit\"{a}t M\"{u}nchen, 
\\
James-Franck-Stra\ss e, 85748 Garching, Germany}
\\[2pt]
{\small  \it 
$^b$
Department of Physics, University of Toyama,
3190 Gofuku, Toyama 930-8555, Japan}
\\[2pt]
{\small  \it 
$^c$
Institute for Theoretical Physics, University of T\"{u}bingen,
Auf der Morgenstelle 14, 72076 T\"{u}bingen, Germany}

\date{\today}

\vskip 1.5cm

\begin{abstract}
Electroweakly interacting stable spin-1 particle in the $(1-10)$~TeV mass range can be a dark matter candidate with rich testability. In particular, one or even two gamma-ray line-like features are expected to be a smoking-gun signature for indirect detection in this scenario. The presence of large Sudakov logarithmic corrections, though,  significantly complicates the theoretical prediction of the gamma-ray spectrum. We resum these corrections at the next-to-leading-log (NLL) accuracy using Soft-Collinear Effective field Theory (SCET). Rather interestingly, we find that the LL- and NLL-resummed endpoint spectra for this model are, up to an overall factor, identical to already existing calculations in the contexts of spin-0 and spin-1/2 (i.e. wino-like) scenarios. We discuss how this non-trivial ``exact universality'' irrespective of DM spin at these accuracies comes about despite the completely different SCET operator bases. Our resummations allow us to reduce the uncertainty, demonstrated in the energy spectrum with distinctive two peaks from annihilations into $\gamma  \gamma, Z  \gamma$ channel and a photon with $Z_2$-even extra heavy neutral boson $Z'$. We discuss the prospect of improving accuracy further, which is crucial for the heavier DM mass region and realistic resolution in future gamma-ray observations.

\end{abstract}
\end{center}
\end{titlepage}

\renewcommand{\thefootnote}{\arabic{footnote}}
\setcounter{footnote}{0}

\tableofcontents

\section{Introduction}

\noindent
We have accumulated evidence of Dark Matter (DM) in various scales through gravitational interactions, e.g. rotation curves of galaxies. Besides, DM may interact with the Standard Model (SM) particles and could be thermalized in the early Universe. Following the freeze-out mechanism, we can predict the DM energy density today as a thermal relic. If the DM-SM interaction is on the order of the (electro)weak interaction of the SM, and if the DM mass is on the TeV scale, the predicted abundance via freeze-out matches the observations of the \emph{Planck} experiment, $\Omega  h^2  =  0.120 \pm 0.001$~\cite{Planck:2018vyg}. This coincidence, commonly dubbed the ``WIMP miracle'', where WIMP stands for weakly interacting massive particle, motivates us to consider DM candidates with electroweak charges. A popular candidate is the so-called ``neutralino'' DM with spin-$1/2$, which is predicted in the Supersymmetry (SUSY) framework. Using this model as a benchmark for more general theories, we list below a set of features that will be particularly relevant in this paper.
\begin{itemize}
\item  DM Particle mass should be the TeV scale to explain the correct thermal relic.
\item  The partners in the multiplet appear with tiny mass splitting ($\sim  {\cal  O}(100)~\mathrm{MeV}$).
\item  All-order corrections (e.g. Sommerfeld enhancement effect, Sudakov logarithms) play a crucial role in achieving precise predictions for phenomenology.
\end{itemize}

In addition to the particle's mass and electroweak multiplicity, its spin is another crucial parameter in the DM theory. Such a building of models has already been systematically studied for spin-$0$ and $1/2$ particles in the past and the resulting models are known as Minimal DM~\cite{Cirelli:2005uq,Cirelli:2007xd,Cirelli:2009uv}.

In the case of spin-$1$ DM, which is the focus of this work, things are a bit more complicated as we need a stabilization mechanism for the massive spin-$1$ spectrum. One direction is the DM theory with extra-dimension~\cite{Flacke:2008ne,Flacke:2017xsv,Maru:2018ocf} where Kaluza-Klein parity stabilizes spin-1 DM candidate. The other possibility is to extend gauge symmetries with exchange symmetry, which is inspired by the mechanism of deconstructing dimension~\cite{Arkani-Hamed:2001kyx,Arkani-Hamed:2001nha}. Thanks to this technique, we can assign a $Z_2$ parity symmetry for the physical spectrum and realize a $Z_2$-odd spin-$1$ spectrum that has electroweak triplet features~\cite{Abe:2020mph} even as the renormalizable model. Using this benchmark model for triplet spin-$1$ DM, we can compare predictions for, e.g., indirect detection among several DM candidates with different spins and discuss the strategy of how to distinguish DM spin in experimental searches.

This paper aims to derive precise indirect-detection predictions for the aforementioned electroweak-triplet spin-$1$ DM candidate. More specifically, we provide state-of-the-art calculations for gamma-ray spectra that will be searched for by current and next-generation Cherenkov telescopes. 
Our work is thus not only crucial for the correct interpretation of the relevant astronomical observations but also for the determination of the spin of the DM particle.

The first attempt for these calculations, e.g. tree-level matching and Sommerfeld enhancement effects, have already been completed, see Ref.~\cite{Abe:2021mry}. However, as has been known for a while, Sudakov-log corrections can be very large for the phenomenology of TeV-scale electroweak-interacting particles, see Ref.~\cite{Hryczuk:2011vi,Cohen:2013ama}. The systematical resummation of such large log corrections can be done in the context of Soft-Collinear Effective Theory (SCET), as the relevant observable cross sections can be mathematically factorized in terms of well-defined renormalization-group-evolution (RGE) properties.

Regarding the DM spin, we can already anticipate that the spin structure here is more complex than those for spin-0 and spin-1/2 DM scenarios~\cite{Baumgart:2014vma,Bauer:2014ula,Baumgart:2018yed,Beneke:2019vhz,Beneke:2020vff};
we need new operators to match the full theory amplitude since spin-$1$ DM pair may form states with higher spins, up to a total spin-$2$ state, to annihilate. Regarding the electroweak features, on the other hand, we can largely recycle the previous results such as the anomalous dimension for the effective operators since these quantities are determined only by SU(2)$_L$ properties of DM. Putting all the pieces together, we can finally present our main calculation: the fully-resummed  endpoint \emph{gamma-ray spectrum at the NLL accuracy for spin-1 DM annihilation}. This final result will be shown in a more complete picture, combining with the continuum gamma-ray spectrum and the second monochromatic gamma-ray line from the annihilation in association with other neutral heavy bosons predicted in our spin-$1$ DM model.

The rest of the paper is organized as follows. In Sec.~\ref{sec:Model}, we introduce our model of electroweak-triplet spin-1 DM and its distinctive indirect-detection signatures in gamma-ray astronomy. In Sec.~\ref{sec:factorization_formula}, we derive factorization formulas for spin-1 DM annihilation cross section in the framework of SCET. We devote a large part to specifying the operators for DM annihilation into electroweak bosons. By matching the Wilson coefficient with full theory amplitudes, we explicitly prove the DM-spin universality of the resummed endpoint spectra at next-to-leading log (NLL) accuracy. In Sec.~\ref{sec:results}, we show our numerical result for gamma-ray energy spectra including the smoking-gun feature of the spin-1 DM model; two separable peaks originating from multiple annihilation channels. We compare our result with analog spin-1/2 DM predictions and discuss detection prospects for CTA. We also discuss how we could further improve the accuracy. Our final discussions and conclusions are given in Sec.~\ref{sec:conclusions}. Throughout this paper we adopt the natural unit system $c  =  \hbar  =  1$.

\section{Model}
\label{sec:Model}

We briefly review the renormalizable model of electroweakly interacting spin-1 DM proposed in Ref.~\cite{Abe:2020mph}. In this model, the electroweak symmetry SU(2)$_L\times$U(1)$_Y$ is extended to SU(2)$_0  \times$SU(2)$_1  \times$SU(2)$_2  \times$U(1)$_Y$ to realize additional massive spin-$1$ spectra. We impose an exchange symmetry between SU(2)$_0$ and SU(2)$_2$ to stabilize a DM candidate, as further explained below.  Table~\ref{tab:model} summarizes the matter content of the model. 
%
\begin{table}[tb]
\centering
\caption{The symmetric structure and charge assignment in this model. The generation indices for the matter fields are implicit.}
\label{tab:model}
 \begin{tabular}{cc|ccccc}\hline
  field    & spin           & SU(3)$_c$ & SU(2)$_0$ & SU(2)$_1$ & SU(2)$_2$ & U(1)$_Y$ \\ \hline \hline
  $q_L$    & $\frac{1}{2}$  &   3   &   1       &    2      &     1     &   $\frac{1}{6}$ \\
  $u_R$    & $\frac{1}{2}$  &   3   &   1       &    1      &     1     &   $\frac{2}{3}$ \\
  $d_R$    & $\frac{1}{2}$  &   3   &   1       &    1      &     1     &  -$\frac{1}{3}$ \\
  $\ell_L$ & $\frac{1}{2}$  &   1   &   1       &    2      &     1     &  -$\frac{1}{2}$ \\
  $e_R$    & $\frac{1}{2}$  &   1   &   1       &    1      &     1     &  -1 \\ \hline
  $H$      & 0              &   1   &   1       &    2      &     1     &   $\frac{1}{2}$ \\
  $\Phi_1$ & 0              &   1   &   2       &    2      &     1     &   0 \\
  $\Phi_2$ & 0              &   1   &   1       &    2      &     2     &   0 \\ \hline \hline
 \end{tabular}
\end{table}
%
In the fermion sector, all fields are charged under SU(2)$_1$, and the hypercharge assignment for each field is identical to the SM. In the scalar sector, we introduce $H$, which corresponds to the SM Higgs, and new scalars $\Phi_1$ and $\Phi_2$ charged under SU(2)$_{0,1,2}$ to break the extended symmetry.

The Lagrangian for the bosonic fields is shown below
\begin{align}
\label{eq:UVfull}
  {\cal  L}
  &=
  -  \frac{1}{4}  \sum_{i=0,1,2}  W_{i  \mu  \nu}^A  W_i^{\mu  \nu  A}
  +  D_\mu  H^\dagger  D^\mu  H
  +  \frac{1}{2}  \sum_{j=1,2}  \tr  \left( D_\mu  \Phi_j^\dagger  D^\mu  \Phi_j \right)
  - V  ( \Phi_1, \Phi_1, H ),
\end{align}
where $V$ denotes the scalar potential, and $A=1,2,3$ denote the SU(2)$_L$ indices. We also introduce the field strength for each SU(2) gauge field 
\begin{align}
  W_{i  \mu  \nu}^A  
  &=  
  \del_\mu  W_{i  \nu}^A  -  \del_\nu  W_{i  \mu}^A  +  g_i  \epsilon^{ABC}  W_{i  \mu}^B  W_{i  \nu}^C.
\end{align} 
The Lagrangian is constructed to be invariant under gauge and exchange symmetry transformations. The gauge transformations for each scalar field are
\begin{align}
  \Phi_1  &\mapsto  U_0  \Phi_1  U_1^\dagger,
  &
  \Phi_2  &\mapsto  U_2  \Phi_1  U_1^\dagger,
\end{align}
where $U_{0,1,2}$ are the gauge transformation matrices for the SU(2)$_{0,1,2}$, respectively. The exchange symmetry transformation is, in turn, defined as 
\begin{align}
\Phi_1  &\mapsto  \Phi_2, 
&
\Phi_2  &\mapsto  \Phi_1,
\label{eq:exchange_spin-0}
\\
W_0  &\mapsto  W_2,
&
W_2  &\mapsto  W_0,
\label{eq:exchange_spin-1}
\end{align}
and requires that the gauge couplings of SU(2)$_0$ and SU(2)$_2$ are the same, $g_0  =  g_2$.

Our setup is such that the $\Phi_j  (j=1,2)$ fields acquire non-zero vacuum expectation values (VEVs) around ${\cal O}  (1)~\mathrm{TeV}$, while the $H$ field does not acquire its VEV at this phase. In particular, we assume the following VEVs
\begin{align}
  \Braket{\Phi_j}  &=  
  \begin{pmatrix}
    \frac{v_\Phi}{\sqrt{2}}  &  0
    \\
    0  &  \frac{v_\Phi}{\sqrt{2}}
  \end{pmatrix},
  &
  \Braket{H}  &=  \bm{0}.
\label{eq:VEV}
\end{align}
Due to the diagonal nature of the VEVs in $\Phi_j (j=1,2)$ there is a remnant SU(2) symmetry defined by $U_0  =  U_1  =  U_2$. This SU(2) symmetry can be regarded as the familiar SU(2)$_L$ in the SM.

Furthermore, this VEV respects the discrete exchange symmetry introduced above, which means that all physical degrees of freedom must have the well-defined $Z_2$ \textit{parity}. For instance, notice that the anti-symmetrized spin-1 state of $W_0$ and $W_2$, which is defined as
\begin{align}
V^A  &\equiv  \frac{W_0^A  -  W_2^A}{\sqrt{2}},
\end{align}
behaves like a $Z_2$-odd field under the exchange-symmetry~\eqref{eq:exchange_spin-1} while all the other spin-$1$ fields are $Z_2$ even. Thus, mass mixings between $V^A$ and other spin-$1$ fields are forbidden, and $V^A$ turns out to be a mass eigenstate of the theory. 

The remaining symmetry breaking can be identified with the one in the SM; the $H$ field acquires the VEV as $H = \left( 0,  v/\sqrt{2} \right)^{\rm  T}$ with $v  \simeq  246~\mathrm{GeV}$ to break SU(2)$_L  \times  $U(1)$_Y$ to U(1)$_{\rm  em}$. As the $H$ field has nothing to do with the exchange symmetry~\eqref{eq:exchange_spin-0}-\eqref{eq:exchange_spin-1}, the $Z_2$ parity symmetry found above remains exact even after the symmetry breaking by $\Braket{H}  \ne  0$.

Our main focus is $V^A$, called \emph{$V$ particles} hereafter. 
After the electroweak symmetry breaking, these fields are interpreted as three U(1)$_{\rm  em}$ eigenstates $(V^0$ and $V^\pm)$, the lightest of which is stable thanks to the $Z_2$ symmetry originating from the exchange symmetry. The radiative correction makes the neutral component lightest, and thus $V^0$ can be a stable spin-1 DM candidate~\cite{Abe:2020mph}.

Let us write down the relevant Lagrangian in the SU(2)$_L$ symmetric phase, i.e. with the VEVs~\eqref{eq:VEV}. This will be convenient for the matching with the SCET in Sec.~\ref{sec:factorization_formula}. In this phase, we find three degenerated mass eigenstates for spin-1 fields by diagonalizing the mass matrix in the symmetry breaking phase as denoted  $W^A, V^A,$ and $W'^A$. The electroweak boson $W^A$ remains massless, while two SU(2)$_L$ triplets ($V^A$ and $W'^A$) acquire the following mass
\begin{align}
  m_V  
  &=
  \frac{g_0  v_\Phi}{2},
  &
  m_{W'}  &=  \frac{\sqrt{g_0^2  +  2  g_1^2}  v_\Phi}{2}.
\end{align}
Notice that $m_{W'}$ is always larger than $m_V$, and $W'^A$ decouples from the DM phenomenology in the limit of $g_0 \ll  g_1$. The gauge bosons for SU(2)$_{0,1,2}$ relate to the mass eigenstates in the SU(2)$_L$ symmetric phase via the following rotation matrix, 
\begin{align}
  \begin{pmatrix}
  W_0^A
  \\
  W_1^A
  \\
  W_2^A
  \end{pmatrix}
  &=
  \begin{pmatrix}
  \frac{1}{g_0}  &  &
  \\
  &  \frac{1}{g_1}  &
  \\
  &  &  \frac{1}{g_0}
  \end{pmatrix}   
  \begin{pmatrix}
  0  &  1  &  1
  \\
  1  &  0  &  0
  \\
  0  &  -1  &  1
  \end{pmatrix}  
  \begin{pmatrix}
  1  &  0  &  -1
  \\
  0  &  1  &  0
  \\
  1  &  0  &  1
  \end{pmatrix}  
  \begin{pmatrix}
  g  &  &
  \\
  &  \frac{g_0}{\sqrt{2}}  &
  \\
  &  &  g
  \end{pmatrix}  
  \begin{pmatrix}
  1  &  0  &  \frac{-  g_0^2  +  2   g_1^2}{2  \sqrt{2}  g_0  g_1}
  \\
  0  &  1  &  0
  \\
  0  &  0  &  - \frac{g_0^2  +  2   g_1^2}{2  \sqrt{2}  g_0  g_1}
  \end{pmatrix}  
  \begin{pmatrix}
  W^A
  \\
  V^A
  \\
  W'^A
  \end{pmatrix}  
  \nn
  \\
  &=
  \begin{pmatrix}
  \frac{g_1}{\sqrt{g_0^2  +  2  g_1^2}}  &  \frac{1}{\sqrt{2}}  &  -  \frac{g_0}{\sqrt{2}  \sqrt{g_0^2  +  2  g_1^2}}
  \\
  \frac{g_0}{\sqrt{g_0^2  +  2  g_1^2}}  &  0  & \frac{\sqrt{2}  g_1}{\sqrt{g_0^2  +  2  g_1^2}}
  \\
  \frac{g_1}{\sqrt{g_0^2  +  2  g_1^2}}  &  -\frac{1}{\sqrt{2}}  &  -  \frac{g_0}{\sqrt{2}  \sqrt{g_0^2  +  2  g_1^2}}
  \end{pmatrix}
  \begin{pmatrix}
  W^A
  \\
  V^A
  \\
  W'^A
  \end{pmatrix},
  \label{eq:rotation}
\end{align}
where we introduce 
\begin{align}
  g  &=  \frac{g_0  g_1}{\sqrt{g_0^2  +  2  g_1^2}},
\end{align}
which corresponds to the SU(2)$_L$ gauge coupling. In the first line of Eq.~\eqref{eq:rotation}, we outline in a factorized form each step in the mass-eigenstate diagonalization: 
(i) absorption of the gauge coupling into fields, 
(ii) decomposition of $Z_2$-even and odd components,
(iii) mass diagonalization of $Z_2$-even state,
(iv) coupling exclusion from gauge fields, and
(v) compensation of kinetic mixing term.\footnote{This factorized form has been derived in the SU(2)$\times$SU(2) scenario~\cite{Pappadopulo:2014qza}. See Sec.~4.1 in the reference.} The final result in the second line can also be directly derived by diagonalizing the mass matrix. In this framework, $V$-particles have two mediators, $W^A$ and $W'^A$, both of which are $Z_2$ even spin-$1$ particles and interact via the non-abelian vector couplings. In the following, we focus on the electroweak bosons $W^A$ as the mediator and neglect $W'^A$, which is much heavier than the electroweak scale and only gives sub-leading effects.

The relevant couplings for annihilations of $V$-particles into the electroweak bosons are expressed below. 
\begin{align}
  {\cal  L}_{\rm  vector}
  &=
  -  \frac{1}{4}  W_{\mu  \nu}^A  W^{\mu  \nu  A}
  -  \frac{1}{4}  \left( D_{[\mu}  V_{\nu]}^A \right)  \left( D^{[\mu}  V^{A  \nu]} \right)
  \nn
  \\
  &
  \quad
  +  \frac{1}{2}  m_V^2  V^A_\mu  V^{A\,\mu}-\frac12g\epsilon^{ABC} W_{\mu\nu}^A 
 V^{B\,\mu}V^{C\,\nu},
  \label{eq:LagSU(2)L}
\end{align}
where 
\begin{align}
  W_{\mu  \nu}^A 
  &=  
  \del_\mu  W_{\nu}^A  -  \del_\nu  W_{\mu}^A  +  g  \epsilon^{ABC}  W_{\mu}^B  W_{\nu}^C,
  \\
  D_{[\mu}  V_{\nu]}^A  
  &\equiv  
  \del_\mu  V_\nu^A  -  \del_\nu  V_\mu^A  
  +  g  \epsilon^{ABC}  \left( W_\mu^B  V_\nu^C  -  W_\nu^B  V_\mu^C \right),
  \\
  m_V^2  
  &=
  \frac{g_0^2  v_\Phi^2}{4}.
\end{align}
The Lagrangian is manifestly SU(2)$_L$ gauge invariant under the gauge transformation for $W^A$ and $V^A$
\begin{align}
  W_\mu^A  T^A
  &\mapsto  
  U  \left( W_\mu^A  T^A  +  i  \del_\mu \right)  U^\dagger,
  \\
  V_\mu^A T^A
  &\mapsto  
  U  \left( V_\mu^A  T^A \right)  U^\dagger,
\end{align}
where we introduce the SU(2)$_L$ symmetry rotation matrix $U  \equiv  \exp  ( i  \alpha^A(x)  T^A )$.

After the symmetry breaking of SU(2)$_L$, the interaction terms in Eq.~\eqref{eq:LagSU(2)L} induce V-particle annihilation into electroweak bosons. One of the most promising strategies is to search for final states including quasi-monochromatic photons (henceforth gamma-ray lines), such as $\gamma  \gamma$, $\gamma  Z$, and $\gamma  Z'$. Here, $Z'$ is the $Z_2$-even heavier neutral boson originating from $W'^A$. In the following, let us use $X$ to denote any multi-particle configuration emitted in the opposite direction of the photon, which is allowed by kinematics and other conservation laws (e.g. $X=W^+W^-,ZH,\ldots$). The line energy in $V^0  V^0  \to  \gamma  +  X$ process is given by
\begin{align}
E_\gamma  =  m_V  -  \frac{m_X^2}{4  m_V}, 
\label{eq:energy_Z'}
\end{align}
where $m_X$ is the invariant mass of $X$ state. Since the photon is massless and $m_Z\ll m_V$, annihilations into $\gamma  \gamma, \gamma  Z$ are nearly degenerated at $E_\gamma  \sim  m_V$, assuming the energy resolution is not enough to distinguish these peaks, $\Delta  E_\gamma  \gtrsim  m_Z^2/(4  m_V^2)$. On the other hand, the $\gamma  Z'$ channel predicts another separable peak within the viable parameter range of mass ratio $m_{Z'}/m_V$, which is a free parameter in this renormalizable spin-1 DM theory. In particular, we focus on $1.02  \lesssim 
 m_{Z'}/m_V  \lesssim  2$ in the following discussion. The lower bound comes from the perturbative unitarity of $g_0$~\cite{Abe:2021mry} while the upper bound comes from the kinematical threshold for $\gamma Z'$.

From Eq.~\eqref{eq:energy_Z'}, the energy resolution required to distinguish the $\gamma  Z'$ peak from the SM peak is $\Delta  E_\gamma  \gtrsim  25\%$. This requirement is expected to be achieved for the ${\cal  O}  (1-10)~\mathrm{TeV}$ region in the future gamma-ray telescope such as Cherenkov Telescope Array (CTA). The prediction of two experimentally distinguishable peaks in the energy spectrum is one of the most characteristic signatures in this model. 
We will further discuss this smoking gun signature when we show our energy spectrum in Sec.~\ref{sec:results}.

To conclude this section, let us briefly state the experimentally observable quantity that we want to predict theoretically. Namely, the gamma-ray flux from DM annihilations in the Galactic halo, which is given by
\begin{equation}
    \Phi=\frac1{8\pi m_V^2}J(\Delta\Omega)\frac{\diff(\sigma v)}{\diff E_\gamma} \ ,
\end{equation}
where $J(\Delta\Omega)$ is the astrophysical $J$-factor as a function of the subtended solid angle $\Delta\Omega$. 
It depends on the mass distribution of the DM ($\rho_{\rm DM}$) in the Galaxy and it is obtained by 
\begin{equation}
    J(\Delta\Omega)=\int_{\Delta\Omega}\diff\Omega\int_{\rm l.o.s.}\diff s \rho_{\rm DM}^2\ ,
\end{equation}
where the second integral denotes the \emph{line of sight} (l.~o.~s.) integral.

The remaining part, which we shall refer to from now on as the \emph{spectrum} $\diff(\sigma v)/\diff E_\gamma$, will be the focus of this work.
This is the velocity-averaged differential cross section for the annihilation of DM particles into gamma-rays ($V^0V^0\to\gamma+X$) with energy $E_\gamma$, times their relative velocity.

\section{Factorization formula}
\label{sec:factorization_formula}

The theoretical prediction of the spectrum is quite complex since the problem involves the interplay between the several characteristic scales listed below;
\begin{itemize}
\item 
the Center-of-Mass (CM) energy for annihilation, $\sqrt s = 2m_V+\mathcal O(v^2)$, where the quantity $v$ is the relative speed of the DM particles\footnote{Typical values of $v$ in our Galaxy are of $\mathcal O(10^{-3})$ and, unless explicitly stated, we assume $v=0$ in this work.} 

\item  
the invariant mass of the (unobservable) additional particles, $m_X$

\item 
the electroweak symmetry breaking scale, $\sim  m_W$

\item  
the de Broglie wavenumber for DM, $m_V v$
\end{itemize}
Fortunately, the parametric hierarchies between these characteristic scales, e.g. $m_W\ll m_V$, allow us to construct an EFT framework in which the annihilation cross section can be expressed in a factored form for each energy scale.
In this context, we can for instance factor out very large contributions coming from nonrelativistic momentum modes linked to the initial DM-pair state. This is done by solving a Schr\"odinger equation with a suitable static potential. See Ref.~\cite{Hisano:2004ds,Arkani-Hamed:2008hhe} and references therein. The procedure is independent of the hard annihilation process, and it effectively resums otherwise uncontrollably large quantum corrections that appear in fixed-order predictions.
On top of these, our factorization formula also resums large double logarithmic effects by the renormalization-group (RG) evolution of several universal and model-specific functions which are associated with the characteristic momentum modes of the annihilation.

\begin{figure}[tb]
    \centering
       \includegraphics[width=0.9\linewidth]{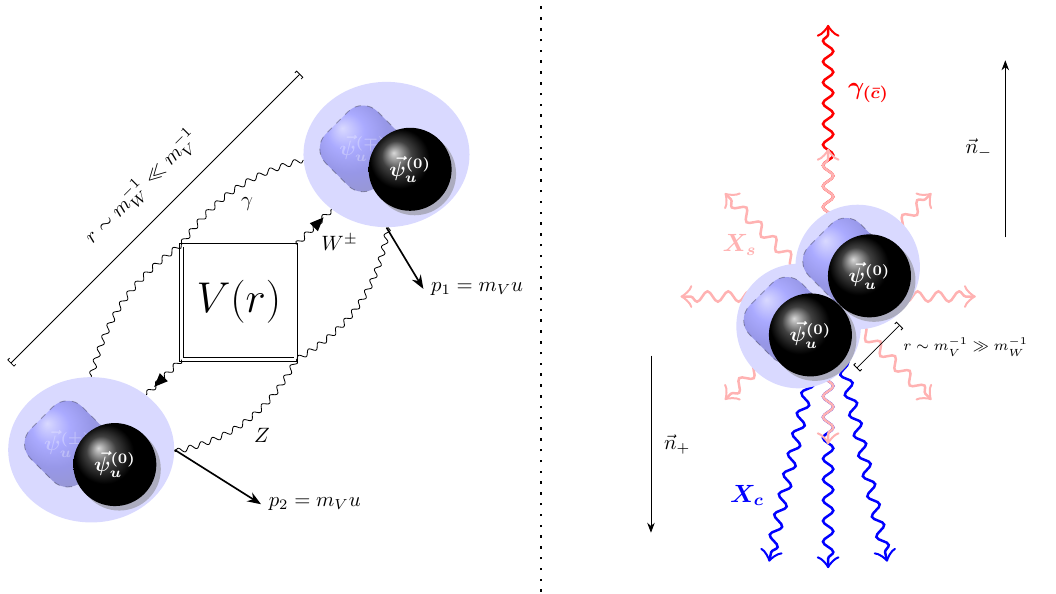}
    \caption{\textit{Left:} 
    initial state non-relativistic dynamics. The DM particles interact via potential $V(r)$ by electroweak bosons and transit into the charged partners. This phenomenon is described in the Sommerfeld enhancement effect, which will be significant for $m_W^{-1}  \ll  m_V^{-1}$.  
    \textit{Right:} final-state dynamics with the anti-collinear (collinear) reference vector, $\vec{n}_-$ ($\vec{n}_+$). 
    }
    \label{fig:kinematics}
\end{figure}
%
To identify the relevant momentum modes of the problem, we start by introducing the reference $4$-vectors to describe momenta for initial- and final-state particles. Figure~\ref{fig:kinematics} schematically shows spin-1 DM annihilation into $\gamma  +  X$, where the left (right) panels show the initial (final) state. First, we introduce the CM reference 4-vectors, $u=(1,0,0,0)^T$, and the initial-state momenta are given by 
\begin{align}
p_1\simeq p_2\simeq m_V u. \qquad  \text{(momentum for initial DM)}
\end{align} 
Besides, the CM momentum reads $p_{\rm CM}=2m_V u$. Second, we introduce the \emph{anti-collinear} light-like reference vector $n_-$.
In the reference frame where the gamma-ray points into the $z$-axis, this vector reads $n_-=(1,0,0,1)^T$. Using this reference vector, the gamma-ray 4-momentum can be written as 
\begin{align}
p_\gamma=E_\gamma n_-.  \qquad  \text{(momentum for anti-collinear photon)}
\end{align}
Lastly, in the aforementioned frame, it will be useful to define the \emph{collinear} reference vector as $n_+=(1,0,0,-1)^T$.
As shown in Fig.~\ref{fig:kinematics} this vector points towards the opposite spatial direction of the gamma-ray. Note that in this nomenclature the sense of ``collinearity'' is defined concerning the unobservable recoiling particles in the $\gamma+X$ final state. Both anti-collinear and collinear vectors are light-like ($n_-^2=n_+^2=0$), and we can express the momentum for $X$ particle as 
\begin{equation}
    p_X=m_Vn_+ + (m_V-E_\gamma)n_-,  \qquad  \text{(momentum for collinear $X$-particle)}
    \label{eq:p_X}
\end{equation}
which is derived from 4-momentum conservation, $p_{\rm  CM}  =  p_\gamma  +  p_X$.

To illustrate this with a concrete example, let us consider the simplest cases in which $X$ (as in $V^0  V^0  \to  \gamma  +  X$) consists of one single body, e.g. $X=\gamma$. The kinematics is trivial in this case: we obtain $p_\gamma=m_V n_-$, $p_X=m_V n_+$. 
The energy of the gamma-ray thus equals the mass of the DM particle, which gives rise to the well-defined monochromatic line in the spectrum originating from $V^0  V^0  \to  \gamma  \gamma$.
This energy value corresponds to the cutoff energy of the spectrum. This can be shown by invoking the reality condition of the invariant mass of the remaining particles, 
\begin{align}
  m_X^2=4m_V(m_V-E_\gamma)\geq0, 
  \label{eq:mXsq}
\end{align}
derived from Eq.~\eqref{eq:energy_Z'}. The reader can also verify this by squaring $p_X$ in Eq.~\eqref{eq:p_X} using $m_X=\sqrt{p_X^2}$ and $n_+\cdot n_-=2$.

As we discussed in the previous section, $Z$ and $Z'$ can be the unobserved partner $X$. Recall that the $\gamma  Z$ channel is indistinguishable from the $\gamma\gamma$ channel while $Z'$ line is separable depending on CTA's energy resolution.

In the following, we will focus our attention on $m_X\ll m_V$, which is the case for $\gamma  \gamma$ and $\gamma  Z$ channels. Since the $m_X\ll m_V$ condition is equivalent to $m_V-E_\gamma\ll m_V$ (see Eq.~\eqref{eq:mXsq}) we see that the decrease in the $m_X/m_V$ ratio translates into reducing the ``distance'' between the measured photon and the cutoff. 
Sudakov logarithmic effects will thus become larger and larger as $m_X$ is decreased. To systematically resum these effects at all orders in perturbation theory, an EFT approach is necessary.
In this context, the non-relativistic dynamics of the problem can be reinterpreted as a quantum-mechanical problem and we treat the large Sudakov logarithmic effects at the endpoint spectrum using SCET methods.
This framework has already been presented in Refs. \cite{Baumgart:2014vma,Bauer:2014ula,Ovanesyan:2014fwa,Beneke:2018ssm,Ovanesyan:2016vkk,Baumgart:2017nsr,Baumgart:2018yed,Beneke:2019vhz,Beneke:2019gtg} for several DM models and assumptions.

We adopt the assumption that $m_X^2$ is parametrically of $\mathcal O(2m_Vm_W)$, as done in Ref. \cite{Beneke:2019vhz}. To understand why this assumption is reasonable, consider the following momentum decomposition of a generic multi-particle final state such as the one depicted in Fig.~\ref{fig:kinematics}: 
\begin{equation}
\label{eq:momenta}
    p_{\rm CM}  =  p_{\gamma_{\bar c}}+P_{X_c}+P_{X_s}\ , 
\end{equation}
where we added the subscript for each momentum; the photon momentum is characterized by the anti-collinear direction. In addition, we divide $p_X$ into two components, $P_{X_c}$ and $P_{X_s}$, which are sums of the particles' momenta created by the hard-collinear ($X_c$) and soft fields ($X_s$), respectively.

In SCET, we expand the physical process by introducing power-counting parameters, denoted by $\lambda$ in the following, and classify regimes in the construction of SCET. For instance, we expand each momentum in terms of $\lambda$ as follows,
\begin{eqnarray}
    p_{\gamma_{\bar c}} &=& E_\gamma n_-\ , \nonumber\\
    P_{X_c} &=& P^{+(0)}_{X_c}n_++P^{\perp(1)}_{X_c}+P^{-(2)}_{X_c}n_-\ ,\nonumber\\
    P_{X_s} &=& P_{X_s}^{+(1)}n_++P_{X_s}^{\perp(1)}+P_{X_s}^{-(1)}n_-\ ,
\end{eqnarray}
where the superscripts indicate the leading scaling order on $\lambda$, i.e. $P^{(n)}_{X}$ is of $\mathcal O(\lambda^n)$, and components perpendicular to both $n_+$ and $n_-$ are indicated by the $\perp$ symbol. Then, Eq. \eqref{eq:momenta} gives $P^{+(0)}_{X_c}=E_\gamma=m_V$ and $P_{X_c}^{\perp(1)}=-P_{X_s}^{\perp(1)}$ such that the invariant mass of $X$ can be written as
\begin{eqnarray}
    m_X^2 = (P_{X_c}+P_{X_s})^2=4P^{+(0)}_{X_c}P^{-(1)}_{X_s}+\mathcal O(\lambda^2)\simeq 4m_Vp_s\ ,
\end{eqnarray}
where we introduced $p_s$ as the typical soft-momentum parameter. To characterize the corrections to the leading-order energy scale, we define the power-counting parameter as $\lambda  \equiv 
 p_s/(2m_V)$. In terms of $m_X$ and $m_V-E_\gamma$, we introduce the following explicit form
\begin{align}
\lambda  &=  \frac{m_X^2}{8m_V^2} 
=\frac{m_V-E_\gamma}{2m_V}. 
\label{eq:lambda_def}
\end{align}

However, the preceding discussion is still incomplete; the first and second expressions of Eq.~\eqref{eq:lambda_def} are characterized by the unobserved invariant mass of $X$ and energy differences at the endpoint, respectively. Neither of them can be determined experimentally for $m_X  \ll  m_V$ and remain unspecified. More complicatedly, there is another momentum scale that plays a major role in our theoretical prediction. 
Namely, the mass scale of the electroweak gauge bosons involves. 
We thus introduce an additional power-counting and its associated typical soft momentum parameters, $\lambda_W=m_W/(2m_V)\ll1$ and $p_s\sim m_W$, respectively. 

Depending on the relationship between $\lambda$ and $\lambda_W$, different factorization setups can be implemented.
Below we briefly summarize the three different strategies that have been put forward in the literature;
\begin{itemize}
    \item {\bf Wide resolution (\boldmath$\lambda\gg\lambda_W$):} besides the obvious $\lambda\ll1$ condition, the $\lambda$ parameter is unconstrained and such that $\lambda\gg\lambda_W=m_W/(2m_V)$. 
    In this regime, re-factorization of the soft modes ($p_s\sim2m_V\lambda $ to $p_s\sim2m_V\lambda_W$) is required. 
    This situation has been considered in Refs. \cite{Baumgart:2017nsr,Baumgart:2018yed}
    
    \item {\bf Intermediate resolution (\boldmath$\lambda\sim\lambda_W$):} $\lambda$ is constrained to be parametrically of $\mathcal O(\lambda_W)$.
    The big advantage of this calculation is that no re-factorization of the soft modes is required, and it has been considered in Refs. \cite{Beneke:2019vhz,Beneke:2019gtg}.
    
    \item {\bf Narrow resolution (\boldmath$\lambda\sim\lambda_W^2\ll\lambda_W$):} $\lambda$ is constrained to be of $\mathcal O(\lambda_W^2)$ so that the real soft states ($X_s$) are composed of ultra-soft photons ($p_s\sim m_W^2/(2m_V)$) whose contributions are power suppressed. See Ref. \cite{Beneke:2019vhz,Beneke:2018ssm,Ovanesyan:2016vkk}. 
\end{itemize}
In general, we need to interpolate these procedures to scan the parameter space of the DM mass. The critical DM mass to separate these regimes is $m_V  \sim  m_X^2/(4  m_W)$, and thus narrow or intermediate resolution is the most appropriate for the $\mathcal O(1)~\mathrm{TeV}$ DM scenarios. As already mentioned, we will consider the intermediate resolution with the corresponding scaling assumptions in the following to make a first step to comprehensively analyze the whole DM mass region.

\subsection{Soft-collinear EFT}
\label{sec:operator_basis}

We start by introducing three spin-1 NR fields $\vec\psi_u^{(0)}$ and $\vec\psi_u^{(\pm)}$, whose kinetic and potential terms in the EFT Lagrangian read 
\begin{align}
\label{eq:pnreft}
    \mathcal L_{\rm pNREFT} &{}={} \vec\psi^{(0)\dagger}\mvdot\left(i\partial_t+\frac{\vec\partial^2}{2m_V}\right)\vec\psi^{(0)}+\sum_{q=\pm}\vec\psi^{(q)\dagger}\mvdot\left(i\partial_t-\delta m+\frac{\vec\partial^2}{2m_V}\right)\vec\psi^{(q)}\nonumber\\
    {}-\frac13&\sum_{\{q_i\}}\int\diff^3\bm{r}\left(\, V^{S=0}_{q_1q_2q_3q_4}(r)\vec\psi^{(q_1)\dagger}(t,\bm{x})\mvdot\vec\psi^{(q_2)\dagger}(t,\bm{x+r})\vec\psi^{(q_3)}(t,\bm{x})\mvdot\vec\psi^{(q_4)}(t,\bm{x+r})\right.\nonumber\\
    {}-\frac12\,{}& V^{S=1}_{q_1q_2q_3q_4}(r)[\vec\psi^{(q_1)\dagger}(t,\bm{x})\times\vec\psi^{(q_2)\dagger}(t,\bm{x+r})]\mvdot[\vec\psi^{(q_3)}(t,\bm{x})\times\vec\psi^{(q_4)}(t,\bm{x+r})]\nonumber\\
    {}-\frac12\,{}& V^{S=2}_{q_1q_2q_3q_4}(r)\left[\vec\psi^{(q_1)\dagger}(t,\bm{x})\mvdot\vec\psi^{(q_3)}(t,\bm{x})\vec\psi^{(q_2)\dagger}(t,\bm{x+r})\mvdot\vec\psi^{(q_4)}(t,\bm{x+r})+\right.\nonumber\\
    {}&{}\left.+\,\vec\psi^{(q_1)\dagger}(t,\bm{x})\mvdot\vec\psi^{(q_4)}(t,\bm{x+r})\vec\psi^{(q_2)\dagger}(t,\bm{x+r})\mvdot\vec\psi^{(q_3)}(t,\bm{x})\right.\nonumber\\
    {}&{}\left.\left.-\,\frac23\vec\psi^{(q_1)\dagger}(t,\bm{x})\mvdot\vec\psi^{(q_2)\dagger}(t,\bm{x+r})\vec\psi^{(q_3)}(t,\bm{x})\mvdot\vec\psi^{(q_4)}(t,\bm{x+r})\right]\right)\ .
\end{align}
Note that this is valid after some field re-definitions that remove the couplings of the NR fields with ultra-soft electromagnetic fields. The potential for DM particles has the same form as other SU(2)$_L$ triplet DM candidates, see e.g. \cite{Beneke:2014gja}.

The annihilation is more naturally described in terms of the unbroken NR fields $\vnr^A_\mu$. The indices $A=1,2,3$ and $\mu=0,1,2,3$ are the (unbroken) SU(2)${}_L$ adjoint and Lorentz indices, respectively.
We use the latter, instead of spatial indices as in Eq. \eqref{eq:pnreft}, for convenience. The translation into the broken-symmetry fields $\vec\psi^{(q)}$ is done by noting that $u\cdot\vnr^A=0$ and
\begin{equation}
    \vec\psi^{(0)} = \vec{\vnr}^3 \quad ,\quad \vec\psi^{(\pm)} = \frac{\vec{\vnr}^1\mp i \vec{\vnr}^2}{\sqrt2}\ . 
\end{equation}
The hard-collinear SCET fields that are relevant for the following discussion are denoted as $\scA^A_{c,\mu}$, $\scA^B_{\bar c,\nu}$ where the subscripts $c$ and $\bar c$ indicate whether the SCET fields are collinear or anti-collinear concerning the observed gamma-ray (see Fig. \ref{fig:kinematics}).

The couplings of SCET fields with the NR DM fields are controlled by an NRDM/SCET operator basis ($\{\mathcal O_{\mJ}^{(S)_{i}}\}$); We classify the operator basis by the total spin ($S=0,1,2$), the total SU(2)$_L$ charge (${\cal  J}  =  0,1,2$), and the label $i=1,2$ to distinguish the $S=2$ operator basises. The effective Lagrangian density expands as
\begin{equation}
\label{eq:Lscet}
    \mathcal L_{\rm int}=\frac1{2m_V}\sum_{S=0}^2\sum_{\mJ=0}^2\sum_i\int\diff s\,\diff t\,\tilde C^{(S)_i}_\mJ(s,t,\mu)\mathcal O_\mJ^{(S)_i}(t,s,\mu)\ ,
\end{equation}
where the first two arguments inside the coefficient functions $\tilde C_\mJ^{(S)_i}(s,t,\mu)$ parameterize integrations along the collinear and anti-collinear directions on which the SCET fields depend, while the third argument $\mu$ is the renormalization-group scale parameter. Possible forms for the operators are specified by imposing suitable symmetry, such as Bose symmetry for both DM fields and electroweak bosons. See Appendix~\ref{sec:operators} for a detailed discussion. The operators to describe DM annihilation signatures are given by 
\begin{eqnarray}
\label{eq:opbasisi}
    \mathcal O^{(0)}_\mJ &=& \tilde\vnr_\alpha^A\eta^{\alpha\beta}_u\tilde\vnr_\beta^B T_\mJ^{ABCD}\tilde\scA^C_{c,\mu}(s n_+)\eta_\perp^{\mu\nu}\tilde\scA^D_{\bar c,\nu}(t n_-)\ ,
    \\
    \mathcal O^{(2)_1}_\mJ &=& \tilde\vnr_\alpha^A\tilde\vnr_\beta^B T_\mJ^{ABCD}(\eta_\perp^{\alpha\mu}\eta_\perp^{\beta\nu}+\eta_\perp^{\alpha\nu}\eta_\perp^{\beta\mu})\tilde\scA^C_{c,\mu}(s n_+)\tilde\scA^D_{\bar c,\nu}(t n_-)-\frac23\,\mathcal O^{(0)}_\mJ\ ,
    \\
    \mathcal O^{(2)_2}_\mJ &=& \tilde\vnr_\alpha^A(n_+-n_-)^\alpha\tilde\vnr_\beta^B(n_+-n_-)^\beta T_\mJ^{ABCD}\tilde\scA^C_{c,\mu}(s n_+)\eta^{\mu\nu}_\perp\tilde\scA^D_{\bar c,\nu}(t n_-)\nonumber\\
    {}&{}&{}\ +\frac43\,\mathcal O^{(0)}_\mJ\ ;
\label{eq:opbasisf}
\end{eqnarray}
where the fields with tildes have not been yet decoupled from the soft electroweak gauge bosons via Wilson line decoupling transformations. See Ref. \cite{Beneke:2019gtg} for a thorough discussion. The Lorentz indices in Eqs.~\eqref{eq:opbasisi}-\eqref{eq:opbasisf} are contracted using the following projectors
\begin{eqnarray}
    \label{eq:metric_ini}
    \eta_u^{\mu\nu} &=& \eta^{\mu\nu}-u^\mu u^\nu\ 
    =  \mathrm{diag} \left( 0,-1,-1,-1 \right),
    \\
    \eta_\perp^{\mu\nu} &=& \eta^{\mu\nu}-\frac{n_+^\mu n_-^\nu+n_-^\mu n_+^\nu}2 \ 
    =  \mathrm{diag} \left( 0,-1,-1,0 \right).
    \label{eq:metric_fin}
\end{eqnarray}
After this projection, the $0$-th component of the contracted vector is switched off to match with the result of full theory in the NR limit. The SU(2)${}_L$ singlet ($\mJ=0$) and quintuplet ($\mJ=2$) structures are given by
\begin{eqnarray}\label{eq:TJ_ini}
    T_{\mJ=0}^{ABCD}&=&\delta^{AB}\delta^{CD}\ ,\\
    T_{\mJ=2}^{ABCD}&=&\delta^{AC}\delta^{BD}+\delta^{AD}\delta^{BC}-\frac23\delta^{AB}\delta^{CD}\ .
    \label{eq:TJ_fin}
\end{eqnarray}
In total there are two spin-0 operator ($\mathcal O^{(0)}_0$ and $\mathcal O^{(0)}_2$) and four spin-2 ones ($\mathcal O^{(2)_1}_0$, $\mathcal O^{(2)_1}_2$, $\mathcal O^{(2)_2}_0$ and $\mathcal O^{(2)_2}_2$).

Notice that an SU(2)$_L$-triplet operator is also allowed from the symmetry argument,
\begin{equation}
    \mathcal O^{(1)}_1 = \tilde\vnr_\alpha^A\tilde\vnr_\beta^B(\eta^{\alpha\mu}_\perp\eta^{\beta\nu}_\perp-\eta^{\alpha\nu}_\perp\eta^{\beta\mu}_\perp)(\delta^{AC}\delta^{BD}-\delta^{AD}\delta^{BC})\tilde\scA^C_{c,\mu}(s n_+)\tilde\scA^D_{\bar c,\nu}(t n_-)\ .
\end{equation}
However, this operator is not relevant to our photon spectrum computation as discussed in detail below.

\subsection{Factorization}
Neglecting power-suppressed corrections, the endpoint photon spectrum is given by
\begin{equation}
\label{eq:factorization}
    \frac{\diff (\sigma v)}{\diff E}=2\sum_{I,J}\sum_{S=0,2}\mathcal S_{IJ}^{S}\frac{\diff (\tilde\sigma v)^{S}_{IJ}}{\diff E}\ ,
\end{equation}
where the annihilation matrices can be factored as follows 
\begin{align}
\label{eq:factorizationII}
    \frac{\diff (\tilde\sigma v)^{S}_{IJ}}{\diff E}={}&{}\frac1{\sqrt2^{\textrm{id}(I)+\textrm{id}(J)}}\frac19\frac1{2\pi m_V}\sum_{\mathcal I,\mathcal J=0,2}\sum_{i,j}\kappa_{ij}^{S}H_{\mathcal I,\mathcal J}^{S;i,j} 
    \nn
    \\
    {}&{}
    \times
    Z_\gamma^{33}
    \int\diff\omega J\left(4m_V(m_V-E-\omega/2)\right)W_{\mathcal I,\mathcal J\,;\,IJ}(\omega)\ .
\end{align}
Notice that the total spin $S$ is conserved if we focus on the leading order effects in the NR limit, where the potential has spherically symmetric form. Therefore, the total spin dependence appearing in the first line has a diagonalized form. For instance, 
\begin{eqnarray}
\kappa^{(S=0)}&=&\eta_{v\,\alpha\beta}\eta_v^{\alpha\beta}\eta_{\perp\,\mu\nu}\eta_\perp^{\mu\nu}=6\ ,
\\
    \kappa^{(S=2)}&=&\left(
    \begin{array}{cc}
        \frac{28}3 & \frac{16}3 \\
        \frac{16}3 & \frac{64}3
    \end{array}
    \right),\ 
\end{eqnarray}
are the kinematic factor and matrix, respectively. These are obtained by spin sums of the relevant contracted Lorentz structures of the operator basis \eqref{eq:opbasisi}-\eqref{eq:opbasisf}; e.g.
\begin{align}
    \kappa_{11}^{S=2}={}&{}\sum_{s_1=1}^3\sum_{s_2=1}^3\varepsilon^{\rm NR}_\alpha(s_1)\varepsilon^{\rm NR}_\beta(s_2)\varepsilon^{\rm NR}_{\alpha'}(s_1) \varepsilon^{\rm NR}_{\beta'}(s_2)\left(\eta_\perp^{\alpha\mu}\eta_\perp^{\beta\nu}+\eta_\perp^{\alpha\nu}\eta_\perp^{\beta\mu}-\frac23\eta_v^{\alpha\beta}\eta_\perp^{\mu\nu}\right)\nonumber\\
    {}&{}\times\left(\eta_{\perp\,\mu}^{\alpha'}\eta_{\perp\,\nu}^{\beta'}+\eta_{\perp\,\nu}^{\alpha'}\eta_{\perp\,\mu}^{\beta'}-\frac23\eta_v^{\alpha'\beta'}\eta_{\perp\,\mu\nu}\right)
    \\
    ={}&{}\frac{28}3. 
\end{align}
We impose matching conditions in each energy scale and evolve these to a common energy scale. In our analysis, we take the common reference scale at the soft energy scale. The reference scale $\mu$ dependence is encoded in each function that is calculated in the framework of SCET. The uncertainty arising from the choice of the common energy scale is found to be negligible in spin-1/2 DM system~\cite{Beneke:2019vhz}. Before introducing the factorized functions on each energy scale, let us comment on the RG treatment for the SU(2)$_L$ coupling. The matching conditions between the full-theory amplitude and the amplitude derived in the SCET are given in Ref.~\cite{Bauer:2014ula}. The gauge coupling is evolved via the RGE that include the dark sector in the full theory amplitude, while the DM is integrated out below the hard scale in the SCET. However, these subtleties are only relevant for higher logarithmic accuracies (NLL', NNLL, etc.),  and we thus neglect them here.

The \emph{hard function} $H_{\mathcal I,\mathcal J}^{S;i,j}$ can be expressed in terms of the Fourier-transformed Wilson coefficients given in Eq. \eqref{eq:Lscet} as
\begin{equation}
    H_{\mathcal I,\mathcal J}^{S;\,i,j}(\mu)=C_{\mathcal I}^{(S)_i}(\mu)C_{\mathcal J}^{(S)_j\,*}(\mu)\ ,
\end{equation}
where the Fourier-transformed Wilson coefficients are defined below
\begin{equation}
    C_{\mathcal I}^{(S)_i}(\mu)\equiv\int\diff s\diff t \,\e^{2im_V(s+t)}\tilde C_{\mathcal I}^{(S)_i}(s,t,\mu)\ .
\end{equation}
We determine these coefficients by the matching prescription discussed in Appendix~\ref{sec:Wilson_coefficients-derivation}. The operators in Eqs.~\eqref{eq:opbasisi}-\eqref{eq:opbasisf} have been constructed in such a way that their RG-running is diagonal, i.e. the Wilson coefficients satisfy
\begin{equation}
     C_{\mJ}^{S,\,i}(\mu)=u_{\mJ}(\mu,\mu_i)C_{\mJ}^{S,\,i}(\mu_i),
\end{equation}
where the RG evolution factors $u_{\mJ}(\mu,\mu_i)$ are obtained by solving 
\begin{equation}
    \mu\frac{\diff u_{\mJ}(\mu,\mu_i)}{\diff\mu}=\Gamma_\mJ(\mu) u_{\mJ}(\mu,\mu_i)\ ,
\end{equation}
with the boundary condition $u_\mJ(\mu_i,\mu_i)=1$. For NLL accuracies, $\Gamma_\mJ(\mu)$ is given by \cite{Beneke:2019vhz}
\begin{equation}
    \Gamma_\mJ(\mu) = \gamma_{\rm cusp}\left(2\ln\frac{4m_V^2}{\mu^2}+\frac{i\pi}2(\mJ(\mJ+1)-4)\right)+2\,\gamma_{\rm adj}+\gamma_V\ ,
\end{equation}
in terms of the following anomalous dimensions
\begin{eqnarray}
\gamma_{\rm cusp}^{\rm 2-loop} &=& \frac{\alpha_2(\mu)}\pi+\frac{\alpha_2(\mu)^2}{2\pi^2}\left(\frac{35}{9}-\frac{\pi^2}3\right)+\ldots\ ,\\
\gamma_{\rm adj}^{\rm 1-loop} &=& -\frac{19}{24\pi}\alpha_2(\mu) +\ldots\ ,\\
\gamma_V^{\rm 1-loop} &=& -\frac1{2\pi}\mJ(\mJ+1)\alpha_2(\mu) +\ldots\ .
\end{eqnarray}

In this order, all anomalous dimensions are universal in the sense that they are independent of the spin of the DM particle; notice that we obtain non-trivial constraints among the anomalous dimensions by requiring the factorized amplitude should be independent of the renormalization scale $\mu$. We can separate the constraint into two parts, $\log$ part and the non-$\log$ part. For the $\log$ part, the pre-factor of $\log  \left( 4  m_V^2/\mu^2\right)$ is uniquely determined by the representation of the electroweak bosons and thus independent of the DM spin~\cite{Bauer:2014ula}. For the non-$\log$ part, we require cancelation of the sum of anomalous dimensions, which are not multiplied by log terms over the energy scale. Through this constraint, the non-$\log$ part of the hard anomalous dimension is expressed by the non-$\log$ part of the (anti-)collinear and soft anomalous dimensions. Since the (anti-)collinear and soft physics are clearly independent of DM spin, the non-$\log$ part of the hard anomalous dimension is also spin independent. This is a brief proof of spin independence for anomalous dimensions.

The Wilson coefficients for this model read  
\begin{eqnarray}
    \label{eq:Wilsonini}
    \left.C_0^{(0)}(\mu_V)\right|_{\rm tree}
    =\frac{8\pi}{3}\alpha_2(\mu_V)\, &,&\,
    \left.C_2^{(0)}(\mu_V)\right|_{\rm tree}=-2\pi\alpha_2(\mu_V)\, ,
    \\
    \left.C_0^{(2)_1}(\mu_V)\right|_{\rm tree}=\frac{16\pi}{3}\alpha_2(\mu_V)\, &,&\, 
    \left.C_2^{(2)_2}(\mu_V)\right|_{\rm tree}=-4\pi\alpha_2(\mu_V)\, ,
    \\
    \left.C_0^{(2)_2}(\mu_V)\right|_{\rm tree}=-\frac{4\pi}3\alpha_2(\mu_V)\, &,&\, 
    \left.C_2^{(2)_2}(\mu_V)\right|_{\rm tree}=\pi\alpha_2(\mu_V)\ .
    \label{eq:Wilsonfin}
\end{eqnarray}
The derivation of these coefficients is given in Appendix~\ref{sec:Wilson_coefficients-derivation}. The spin-1 Wilson coefficient accidentally vanishes at this accuracy, which is not rejected from the symmetric argument (see Appendix~\ref{sec:operators}). In any case, the $S=1$ contribution is irrelevant in the nonrelativistic limit; notice that the initial state is limited for two neutral components ($V^0  V^0$) to induce annihilation signature today. As the initial state is composed of two identical spin-1 particles, the Landau-Yang selection rule forbids annihilation into $S=1$ state~\cite{Landau:1948kw,Yang:1950rg}.

The NLL resummation of \emph{jet function} $J(p^2)$ participating in the factorization formula is done as described in Ref. \cite{Beneke:2019vhz}. 
At LL accuracy it reads ($\gamma_E=0.577\ldots$ is the Euler gamma):
\begin{equation}
    \left.J(p^2;\mu_{\rm EW})\right|_{\rm LL}=\exp\left(\frac{\alpha(\mu_J)}\pi\ln^2\frac{\mu_J^2}{\mu_{\rm EW}^2}\right)\frac1{\Gamma(\eta)}\frac1{p^2}\left(\frac{p^2}{\e^{\gamma_E}\mu_J^2}\right)^\eta
\, ,\quad \eta=\frac{\alpha(\mu_J)}\pi\ln\frac{\mu_J^2}{\mu_{\rm EW}^2} \ .
\end{equation}

In our choice of common scale to be soft scale, the \emph{photon jet functions} and the \emph{soft functions} remain unresummed.
These are namely given by $Z_\gamma = \sin^2\theta_W(\mu_{\rm EW})$ and
\begin{equation}
    W_{\mathcal I,\mathcal J\,;\,IJ}(\omega)=D_{\mathcal I;\,I}D_{\mathcal J;\,J}^*\delta(\omega)\ ,
\end{equation}
where the coefficients $D_{\mathcal I;\,I}$ are given by
\begin{align}
    \left.D_{0;(00)}\right|_{\rm tree}={}&{} \left.D_{0;(+-)}\right|_{\rm tree}=1 \ ,\nonumber\\
    \left.D_{2;(00)}\right|_{\rm tree}=\frac43\ ,{}&{} \left.D_{2;(+-)}\right|_{\rm tree}=-\frac23 \ .
\end{align}

\subsection{DM-spin independence of LL/NLL resumation}

Here we discuss the DM spin independence in the resummed gamma-ray spectrum. More concretely, we explain why (up to an overall factor) our spin-1 DM resummed computation of the endpoint gamma-ray spectrum is identical to the already known calculations for spin-0~\cite{Bauer:2014ula} and spin-1/2 (e.g. \cite{Beneke:2019vhz}) DM candidates for LL/NLL accuracies.

This universality has already been discussed in Ref.~\cite{Bauer:2014ula}. Namely, they find universal features that are independent of DM spin in the resummation. The proof of this at the LL accuracy is almost trivial using the SCET machinery, which is one of the greatest advantages of the EFT approach. More specifically, the pre-factor of the LL terms in the RGEs is independent of both spin and the SU(2)$_L$ representation of DM. Since we fix the SU(2)$_L$ multiplicity to be a triplet in our comparisons between DM spins (scalar, fermion, vector), resummation of the Sudakov log part in RGE gives the universal result at LL.

Less trivially, we find a similar result for the NLL accuracy calculation, namely, \emph{the resummed gamma-ray spectra are (up to normalization) spin-independent and the same at the LL and NLL accuracies.} To argue why this is the case, we start by noting that the LL/NLL resummed hard function obeys the following property\footnote{NLL superscripts are used for concreteness. Results hold also for LL accuracies.}
\begin{equation}
    \left.H_{\mathcal I,\mathcal J}^{S;\,i,j}(\mu)\right|_\textrm{NLL} =   h^S_{ij}  \mathcal H_{\mathcal I,\mathcal J}^\textrm{NLL}(\mu) \ , 
\end{equation}
where
\begin{align}    
    h^{(S=0)}&=1 \ ,
    &
    h^{(S=2)}&=
        \begin{pmatrix}
            4 & -1 \\
            -1 & \frac14
        \end{pmatrix}\ ,
\end{align}
and $\mathcal H_{\mathcal I,\mathcal J}^\textrm{NLL}(\mu)$ reads
\begin{equation}
    \mathcal H^\textrm{NLL}(\mu)=
        \begin{pmatrix}
            \frac{64\pi^2\hat\alpha_2^2(\mu_H)}9|u_0(\mu,\mu_V)|^2 & -\frac{16\pi^2\hat\alpha_2^2(\mu_H)}3u_0(\mu,\mu_V)u_2^*(\mu,\mu_V)\\
            -\frac{16\pi^2\hat\alpha_2^2(\mu_H)}3u_0^*(\mu,\mu_V)u_2(\mu,\mu_V) & 4\pi^2\hat\alpha_2^2(\mu_H)|u_2(\mu,\mu_V)|^2
        \end{pmatrix} \ .
\end{equation}
Notably, this is the same (NLL-resummed) hard function that one obtains the factorization formula for the wino model.
The reason why this happens is because the SU(2)$_L$ gauge index structure of tree-level amplitude is identical to that of the spin-0 and spin-1/2 cases. Namely, the total amplitude is proportional to the anti-commutator of SU(2)$_L$ generators,
\begin{align}
\{T^A,T^B\}_{CD}=-(\delta_{AC}\delta_{BD}+\delta_{AD}\delta_{BC}-2\delta_{AB}\delta_{CD}),
\end{align}
where the second expression is for the triplet representation. See Eq.~\eqref{eq:Mtot} for the full amplitude.

In the end, we obtain the a simple relation; the NLL-resummed annihilation matrix reads
\begin{eqnarray}
\nonumber
    \left.\frac{\diff (\tilde\sigma v)^{S}_{IJ}}{\diff E}\right|_\textrm{NLL}&=&f^{(0)}_S\frac1{\sqrt2^{\textrm{id}(I)+\textrm{id}(J)}}\frac14\frac1{2\pi m_V}\kappa^\textrm{wino}\sum_{\mathcal I,\mathcal J=0,2}\mathcal H_{\mathcal I,\mathcal J}^\textrm{NLL}(\mu)\\
\nonumber
    {}&{}&{} \left.Z_\gamma^{33}\right|_\textrm{NLL}
    \int\diff\omega J^\textrm{NLL}\left(4m_V(m_V-E-\omega/2)\right)W_{\mathcal I,\mathcal J\,;\,IJ}^\textrm{NLL}(\omega) \\
    {} &=& {} f^{(0)}_S 
    \left.\frac{\diff (\tilde\sigma v)_{IJ}^\textrm{wino}}{\diff E}\right|_\textrm{NLL}\ ,
\end{eqnarray}
where $\kappa^\textrm{wino}=4$ \cite{Beneke:2019vhz} and 
\begin{align}
    f^{(0)}_S&=\frac19\sum_{i,j}\kappa_{ij}^{S}h_{ij}^{S}=\frac23\ , 
    &
    &(S=0) 
    \\
    f_{S=2}^{(0)}&=\frac{32}9 \ .
    &
    &(S=2) \ 
\end{align}
Since the $S=2$ Sommerfeld factors are identical to those for $S=0$, the full resummed spectrum is proportional to the corresponding wino spectrum:
\begin{equation}
    \left.\frac{\diff (\sigma v)}{\diff E}\right|_\textrm{NLL} = \left.2\mathcal S_{IJ}\left((f_{S=0}^{(0)}+f_{S=2}^{(0)})\frac{\diff (\sigma v)^\textrm{wino}_{IJ}}{\diff E}\right)\right|_\textrm{NLL}=\frac{38}9\left.\frac{\diff (\sigma v)}{\diff E}\right|^\textrm{wino}_\textrm{NLL}\ .
\end{equation}
This relation also holds at the tree-level without Sudakov log resummation since the resummation effect is factorized from the tree-level hard physics, and the same as those for the other DM spins.

To conclude this section let us briefly summarize what we just did. First, we were able to prove the previously advertised DM-spin ``universality''~\cite{Bauer:2014ula} using a suitable SCET for the spin-1 DM model~\cite{Abe:2020mph}, even at NLL accuracy. This universality is obtained because the SU(2)$_L$ structure that appears at the tree-level amplitudes is the same for all DM spins. Specifically speaking, the spin features decouple from the DM process at the leading order of the NR limit.

Second, we also note that this universality will be violated above the NLL accuracy. For spin-0 and 1/2, the ${\cal  O}  (g_2^2)$ terms in Wilson coefficients appear only for the operator that is proportional to $\{T^A, T^B\}_{C D}$, while the ${\cal  O}  (g_2^4)$ corrections appear in a non-universal ways, i.e. model-dependent linear combinations of operators for ${\cal  J}  =  0$ and $2$. Therefore, the spectral shape will no longer be universal, since the RGEs depend on the total SU(2)$_L$ charge ${\cal  J}$ via anomalous dimension. For spin-1 DM, we also expect the higher order corrections to appear in a non-universal way, and thus we need to revisit all the calculations. We postpone the analysis above the NLL accuracy for our future work and show the numerical results including all the features of spin-1 DM in the succeeding section.

\section{Numerical results}
\label{sec:results}

In previous sections we argued that our LL/NLL predictions for the spin-1 DM model's endpoint gamma-ray spectrum differ with respect to the wino (spin-$1/2$) case by an overall factor of $38/9$. 
Moreover, this property also holds in the continuum part of the spectrum as we show in the appendices. Therefore, the numerical impact of the LL/NLL resummation is identical to the one that was already provided in Ref. \cite{Beneke:2019vhz} for the pure-wino model, e.g. Sudakov-log effects at the endpoint spectrum suppress the signal by an $\mathcal O(1)$ factor for $m_V\gtrsim\mathcal O$(TeV). Since the discussion provided there is quite comprehensive, we refer the interested reader to that paper.
We will thus limit ourselves to translating these already existing wino-DM studies to our vector-DM model, and pay special attention to the discussion of theoretical uncertainties at the endpoint spectrum taking into account instrumental effects. We also include the line-like feature from the $V^0  V^0  \to\gamma Z'$ process that is specific to the spin-1 DM model.

All numerical evaluations reported in this paper have been obtained using the same input parameters and renormalization schemes as in Ref. \cite{Beneke:2019vhz} (see Sec. 4 there). In particular, for the computation of the Sommerfeld factors we use on-shell renormalization parameters, e.g. $\alpha_{\rm OS}=1/128.943$, $m_Z=91.1876$~GeV and $m_W=80.385$~GeV. For the computation of the resummed annihilation matrix elements \eqref{eq:factorizationII} we consider running parameters in the $\overline{\rm MS}$ scheme instead, e.g. $\hat\alpha_2(m_Z)=\hat g_2^2(m_Z)/(4\pi)=0.335664$ and $\hat s_W^2(m_Z)=0.232486$, where $\hat g_2(m_Z)$ and $\hat s_W$ are, respectively, the (running) SU(2)${}_L$ coupling of the electroweak theory the sine of the Weinberg angle at $\mu=m_Z$.

\subsection{Recycling prescription for LL/NLL resummations}

In Sec. \ref{sec:factorization_formula}, we discussed the similarities between our resummed annihilation matrix elements in Eq.~\eqref{eq:factorizationII} and their corresponding pure-wino counterparts from Ref. \cite{Beneke:2019vhz}.  
Namely, at LL and NLL accuracies, both formulas yield identical numerical results up to an overall factor of $38/9$ when the $S=0,2$ components in Eq.~\eqref{eq:factorizationII} are added together. 

A similar situation occurs with the NR potentials and their associated Sommerfeld factors: both $S=0$ and $S=2$ potentials in our vector-DM model are identical to the one for winos (see Ref.~\cite{Abe:2021mry}). The only source of numerical differences is the mass splitting parameter $\delta m_V= m_{V^\pm}-m_{V^0}$ which in our case amounts to
\begin{equation}
\label{eq:msplit}
    \delta m_V=168~\textrm{MeV} \ , 
\end{equation}
while for the wino case it is given by $\delta m_\chi^{\rm wino}=164.1$~MeV \cite{Beneke:2018ssm}. The former has been obtained at 1-loop in Ref. \cite{Abe:2021mry} while the latter is known at two loops~\cite{Yamada:2009ve,Ibe:2012sx,McKay:2017xlc}. Even though the numerical difference between these two is rather small, the Sommerfeld factors for near-resonance values of the mass parameter\footnote{In the wino model the Sommerfeld enhancement becomes resonant when the mass parameter equals $2.42$~TeV, $9.36$~TeV, etc. \cite{Beneke:2019qaa}} can be very different. However, away from the resonant regions the differences are negligible.

We may, therefore, ``recycle'' existing LL/NLL resummed calculations for the wino model, e.g. \texttt{DM$\gamma$Spec} \cite{Beneke:2022eci} to obtain the endpoint plus continuum photon spectrum of the vector-DM candidate studied here. As long as the mass parameter $m_V$ is away from a Sommerfeld resonance, the prescription is trivial: \emph{multiply everything by $38/9$}.

Nonetheless, the numerical results that we report here are not obtained using this ``trick'' but we rather re-evaluate all terms occurring in Eqs. \eqref{eq:factorization}-\eqref{eq:factorizationII} using the input parameters stated above (and Ref. \cite{Beneke:2019vhz}) and \eqref{eq:msplit}. We certainly validated our codes with the authors' previous works (e.g. \cite{Beneke:2019vhz,Beneke:2022eci}). 
It should also be noted that the NLO correction of the potential for our model has not yet been computed. Its effect on the spectrum is expected to be sizable in the near-resonance regions as it is the case for wino DM \cite{Beneke:2019qaa,Beneke:2020vff}.

\subsection{The continuum and the \boldmath$Z'$ line}

While the focus of this work is the resummation of Sudakov-log effects at the endpoint of the photon spectrum from our spin-1 DM model, we provide in this section a more complete picture including the continuum and the $Z'$ line parts of the spectrum. The latter is a distinctive feature for this particular model, since $Z'$ should be included to realize a renormalizable model of electroweakly spin-1 DM. 

Let us start by modeling the continuum. This is obtained by matching fixed-order calculations with parton showers. In the appendix, we show that at Born level, all 2-to-2 annihilation matrices, where the final states can be $W^+W^-$, $ZZ$, $\gamma Z$, and $\gamma\gamma$ are, up to a $38/9$ factor, identical to the corresponding annihilation matrices in the wino case.
The continuum spectrum for our model can thus be obtained by multiplying the corresponding continuum part of the photon spectrum for the wino by this $38/9$ factor. As a neat consequence, the smooth matching that was observed in Ref. \cite{Beneke:2022eci} between the endpoint resummations and the continuum part for the wino model is also apparent in our model.
Moreover, the ``recycling prescription'' discussed above can be extended beyond the endpoint all the way down to the continuum, e.g. the complete photon spectrum for wino DM that is generated by \texttt{DM$\gamma$Spec} can be used for our model in the non-resonant mass regions. In this work, however, we obtain the continuum spectra using our own Sommerfeld codes and the VINCIA-based \cite{Fischer:2016vfv} automated tool \texttt{CosmiXs} \cite{Arina:2023eic}.

Our discussion so far seems to indicate that the photon spectrum for our spin-1 DM model is identical to the corresponding spectrum for the wino up to an overall factor. However, our model predicts the existence of further broken gauge bosons ($W'^\pm$ and $Z'$) both of which are heavier than the DM $V^{0}$. If the mass of the neutral $Z'$ is smaller than the center of mass energy of the annihilation $2m_V$, then the $V^0  V^0  \to\gamma Z'$ process is kinematically possible and at first approximation can be described as a monochromatic line (henceforth $Z'$ line):
\begin{equation}
\label{eq:zpline}
    \left.\frac{\diff(\sigma v)}{\diff E_\gamma}\right|_{\gamma Z'}=(\sigma v)_{\gamma Z'}\delta\left(E_\gamma-E_\gamma^{Z'}\right)\ ,
\end{equation}
where \cite{Abe:2021mry}
\begin{equation}
    (\sigma v)_{\gamma Z'}=\frac{\pi\hat\alpha_2^2\hat s_W^2}{36\,m_V^2}\left(76+4\frac{m_{Z'}^2}{m_V^2}+\frac{m_{Z'}^4}{m_V^4}\right)\frac{4m_V^2-m_{Z'}^2}{m_{Z'}^2-m_V^2}\ ,
\end{equation}
and the energy of the $Z'$ line is given by \[E_\gamma^{Z'}=m_V-\frac{m_{Z'}^2}{4\,m_V}\ .\] This feature is a distinctive aspect of our model and is not present in the wino case.

\subsection{Theoretical uncertainties and the full spectrum}
\begin{figure}[t!]
    \centering
        \includegraphics[width=.95\linewidth]{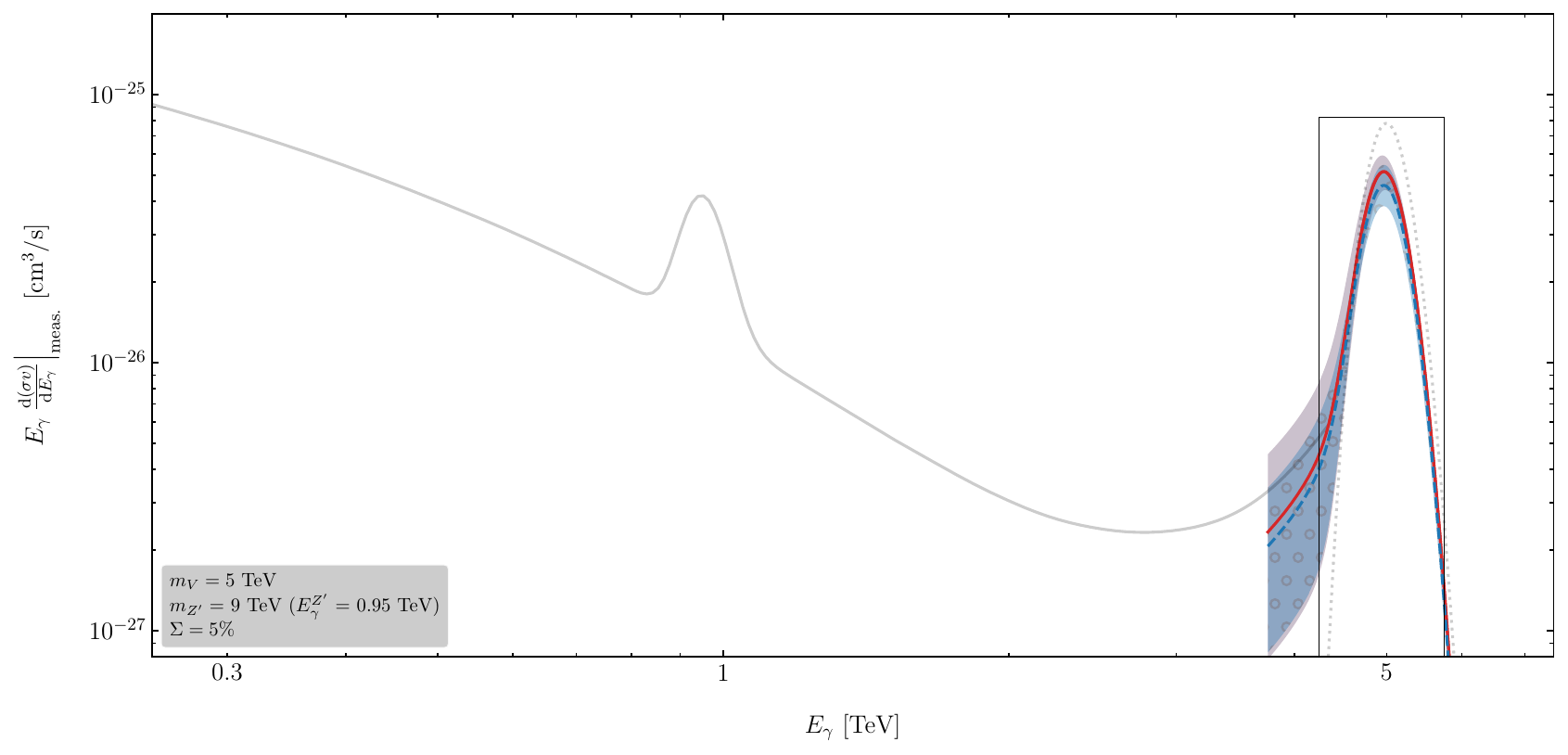}
        \includegraphics[width=0.48\linewidth]{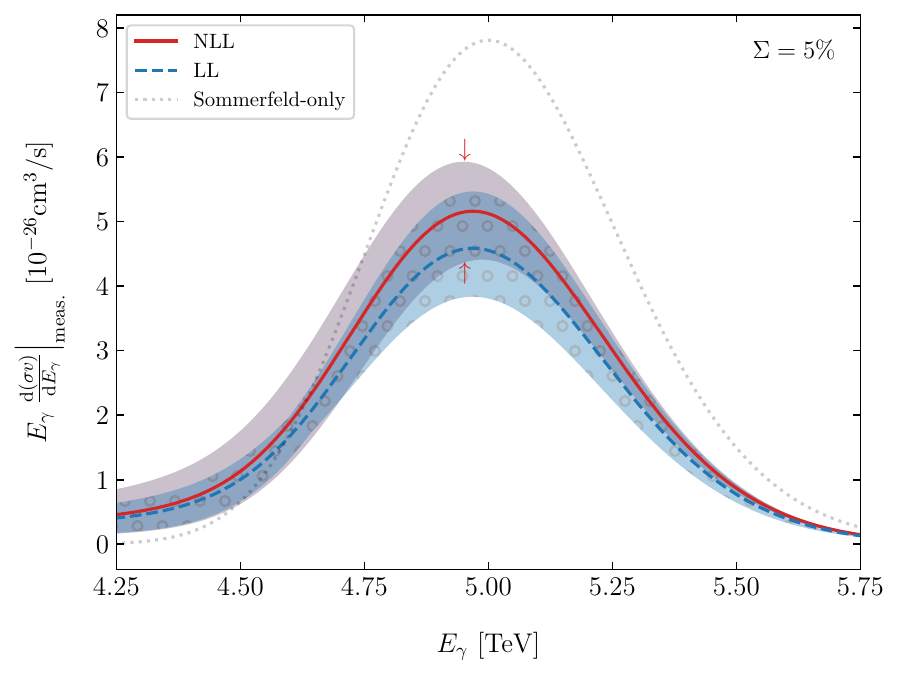}\hspace{10pt}
        \includegraphics[width=0.486\linewidth]{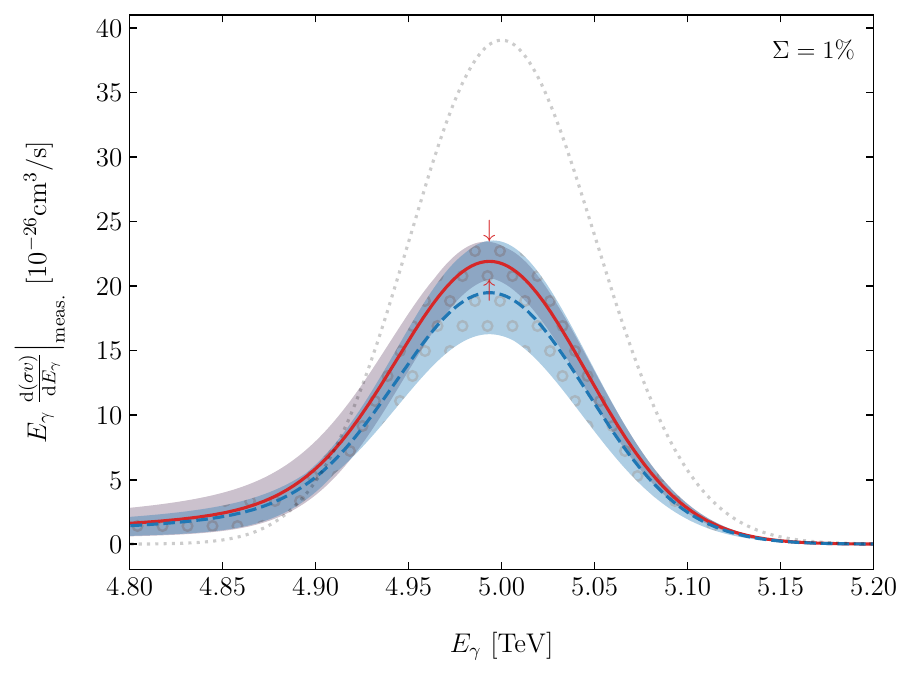}
    \caption{Upper: Full gamma-ray spectrum for our benchmark $5$~TeV spin-1 DM model including instrument-resolution effects (the plotted $\diff(\sigma v)/\diff E_\gamma^{\rm meas.}$ is obtained using Eqs.~\eqref{eq:IRF}- \eqref{eq:PDF} where $\diff(\sigma v)/\diff E_\gamma^{\rm true}$ at the endpoint is given by \eqref{eq:factorization}).}
    \label{fig:diffspec}
\end{figure}

In Fig.~\ref{fig:diffspec}, we show a benchmark example in which our formula is valid. In this model, the DM mass is of $5$~TeV while the $Z'$ boson has a $9$~TeV mass. The $Z'$-line energy amounts to approximately $0.95$~TeV in this setup.
%
\noindent

The width of the two gamma-ray lines in Fig.~\ref{fig:diffspec} is not the natural one\footnote{If instrumental effects were negligible, which corresponds to ideal-detector situation, the $Z'$ line should be treated as a Breit-Wigner resonance with a natural width given by the $Z'$ boson's decay rate.}, but rather we model the instrument response function as a Gaussian distribution with an energy-dependent width, so that the measured spectrum is given by
\begin{equation}
\label{eq:IRF}
    \Phi_{\rm meas.}(E_\gamma)=\int\diff E_\gamma^{\rm true}\mathcal G_\Sigma(E_\gamma,E_\gamma^{\rm true})\Phi_{\rm true}(E_\gamma^{\rm true})\ ,
\end{equation}
where
\begin{equation}
\label{eq:PDF}
    \mathcal G_\Sigma(E_\gamma,E'_\gamma)=\frac1{\sqrt{2\pi}\Sigma\times E_\gamma}\e^{-\frac{(E_\gamma-E'_\gamma)^2}{2\,\Sigma^2E_\gamma^2}}\ .
\end{equation}

The variable $\Sigma$ can be understood as the relative energy resolution of a Cherenkov telescope with a perfectly Gaussian response function, i.e. $\Sigma=0.1$ would correspond to an energy resolution of $10$~\% at the given energy $E_\gamma$. 
Typical energy resolutions for energies $\gtrsim\mathcal O$(TeV) are, of course, much larger than the natural widths of the $Z'$- and $Z$-boson resonances and we can neglect them, which justifies Eq.~\eqref{eq:zpline}.

Concretely, we consider the optimistic but feasible value of $=5$~\% for $\Sigma$ in the main plot (upper panel) of Fig.~\ref{fig:diffspec}. 
In the lower panels, we zoomed in the endpoint part for the same value for $\Sigma$ (lower left panel) and considered the more unrealistic $\Sigma=1$~\% case in the lower right panel as a demonstration. 

Theoretical uncertainties are shown in all plots as shaded bands following the color coding of the several logarithmic accuracies that we consider here (i.e. LL in blue and dashed lines and NLL in red and solid lines). 
Faint dotted lines correspond to ``na\"{\i}ve calculations,'' where Sudakov-log effects are not included. As customary, the bands are obtained by varying in factors of 2 and 1/2 the several virtualities 
($\mu_H  \sim  2m_V$, $\mu_J  \sim  \sqrt{2m_Vm_W}$, $\mu_S\sim m_W$) and rapidity-regulator ($\nu_{H_J}\sim 2m_V$, $\nu_{S_J}\sim m_W$) mass-scale parameters of the hard, jet and soft functions participating in Eq. \eqref{eq:factorization}. Note also that unless $m_{Z'}\simeq m_V$ accidentally, these uncertainties are insensitive to the $Z'$-boson's line phenomenology.

In Ref. \cite{Beneke:2019vhz} a thorough exploration of these uncertainties has already been carried out (see e.g. Fig. 3 in that reference). 
In particular, the LL and NLL results shown in that figure are identical to the ones we obtain here if we consider exactly the same input parameters (and charged-to-neutral particle mass splitting) as in that reference.

We present here (see lower panels of Fig. \ref{fig:diffspec}) for the first time a discussion of the scale variations in predictions that include instrument-resolution effects. Concretely, we can see that uncertainty of the NLL computation at the maximum-flux energy (see red arrows in the plots) goes from 30.25\% for the experimentally accessible (but still theoretically valid if marginally) $\Sigma=5$~\% down to 13.30\% for the theoretically more appropriate (but less feasible experimentally) $\Sigma=1$~\%. This uncertainty reduction can be explained by noticing that our factorization formula \eqref{eq:factorization} is applicable for $\Sigma\sim m_W/(2m_V)$, which for $m_V=5$~TeV is precisely $\simeq1$\%. 

The preceding discussion outlines a clear path forward. Specifically, we are able to enhance the logarithmic accuracy by including loop corrections 
as done in e.g. Ref.~\cite{Beneke:2019vhz}, is crucial for the proper interpretation of gamma-ray line searches below a few TeV. In such cases, the factorization formulas presented here for intermediate (and narrow) energy resolution regimes are well suited. The calculation beyond the NLL, however, exhibit non-universal feature for DM spin, as found in spin-0 and 1/2 cases. Therefore, we must revisit all the computation for spin-1 DM model.

Conversely, for $\mathcal{O}(10$~TeV) spin-1 DM we need to extend ``wide energy resolution'' resummation schemes such as the ones developed in Refs. \cite{Baumgart:2017nsr,Baumgart:2014vma} and apply them accordingly to our spin-1 DM model. Fortunately, given the larger hierarchies inherent in heavy-DM scenarios such extensions at the NLL accuracy are quite robust as shown in those references. Furthermore, a dedicated study on situations in which the splitting between the $Z'$ boson mass and the DM is small is necessary for full explorations of the model's parameter space.

\section{Conclusions}
\label{sec:conclusions}

In this paper, we provide state-of-the-art theoretical predictions for multi-line gamma-ray spectra arising from electroweakly interacting spin-1 
 DM annihilation in the local Universe. We account for large non-relativistic and Sudakov-log effects by employing standard NREFT and SCET methods. We also describe the continuum part of the spectrum using standard parton-shower algorithms.

Besides obtaining robust predictions with NLL theoretical uncertainties of $\mathcal O(10\%)$,  surprisingly, a DM particle spin ``universality law'' was shown. Namely, at both LL- and NLL-accuracies and for the LO static potential, we showed that the endpoint and continuum spectrum of our spin-1 DM particle model is up to an overall factor of $38/9$ identical to that of the spin-1/2 wino model (and its spin-0 counterpart).

The existence of a distinguishable $Z'$ line feature not present in its fermionic and scalar counterpart is what can be used to experimentally discriminate the spin-1 theory studied here. We thus emphasized situations in which such a line is present in the spectrum.

Future work will focus on promoting our calculation to NLL'/NNLL. The main difference is that these calculations include loop corrections, and their impact is stronger when the DM is of $\mathcal O(1$~TeV) or lighter. For heavier DM it is more important to consider factorization schemes for \emph{wider} energy resolutions (e.g. \cite{Baumgart:2017nsr}) than the scheme (intermediate energy resolution) that was considered here.

\section*{Acknowledgments}

\noindent
MF thanks Tomohiro Abe and Stefan Lederer for the useful discussions in the final stage of this project. MF also thanks University of T\"{u}bingen for hospitality during her visit. 
The authors thank the JSPS Core-to-Core Program (grant number: JPJSCCA20200002), where this work was initiated during the ``The 2nd DMNet International Symposium''. We also acknowledge the Mainz Institute for Theoretical Physics (MITP) of the Cluster of Excellence PRISMA$^+$ (Project ID 390831469), for enabling us to discuss this project. This work of MF was supported by the Collaborative Research Center SFB1258 and by the Deutsche Forschungsgemeinschaft (DFG, German Research Foundation) under Germany’s Excellence Strategy - EXC-2094 - 390783311.

\appendix

\section{SCET Operators}
\label{sec:operators}

We give symmetry arguments to specify the form of EFT operators. The effective operators have the following form 
\begin{align}
{\cal  O}_{\cal  J}^{(S,i)}  =  \tilde{\Upsilon}_\alpha^A  \tilde{\Upsilon}_\beta^B  \tilde{\cal  A}_{\perp  c,  \mu}^C  \tilde{\cal  A}_{\perp  \bar{c},  \nu}^D  U^{\alpha  \beta  \mu  \nu}_{(S)}  T_{\cal  J}^{ABCD}, 
\end{align}
where we introduce $T_{\cal  J}$ and $U_{(S)}$ as the tensors for SU(2)$_L$ and Lorentz indices, respectively. We focus on SU(2)$_L$ gauge fields since DM does not have hypercharge. The initial DM has SU(2)$_L$ triplet charge, and we can form total SU(2)$_L$ charge for ${\cal  J}  =  0,1,2$. Explicit form of $T_{\cal  J}$ is given in Eq.~\eqref{eq:TJ_ini}-\eqref{eq:TJ_fin} for ${\cal  J}  =  0,2$ and 
\begin{align}
T_{{\cal  J}  =  1}  =  \delta_{AC}  \delta_{BD}  -  \delta_{AD}  \delta_{BC},
\end{align}
for ${\cal  J}  =  1$. The building blocks for $U_{(S)}$ are listed as $\{ \eta^{\mu  \nu}, \epsilon^{\mu  \nu  \rho  \sigma}, 
n_\pm^\mu \}$.

We classify possible forms for operators into the total spin $S$. In the NR limit, the $0$-th component of a vector particle is suppressed, and we can focus on SO(3) vector $\tilde{\Upsilon}_k  (k=1,2,3)$ to calculate the LO contribution. For a pair of spin-1 DM to form the $S  =  0$ state, we have $\tilde{\Upsilon}_k  \tilde{\Upsilon}_\ell  \delta_{k  \ell}$, which corresponds to $\tilde{\Upsilon}_\alpha  \tilde{\Upsilon}_\beta  \eta_u^{\alpha  \beta}$ for the Lorentz invariant form. The possible structure of Lorentz tensor, therefore, are listed below
\begin{align}
U_{(S=0)}^{\alpha  \beta  \mu  \nu}  
&=  
\left\{
\begin{array}{l}
\eta_u^{\alpha  \beta}  \eta_\perp^{\mu  \nu},
\\
\eta_u^{\alpha  \beta}  \epsilon^{\mu  \nu  \rho  \sigma}  n_{+  \rho}  n_{-  \sigma}.
\end{array}
\right.
\end{align}
For $S=1$, we have $\tilde{\Upsilon}_k  \tilde{\Upsilon}_{\ell}  \epsilon^{k  \ell  m}$, which corresponds to
$\tilde{\Upsilon}_\alpha  \tilde{\Upsilon}_\beta  (n_+  +  n_-)_\rho  \epsilon^{\alpha  \beta  \rho  \sigma}$ as the Lorentz covariant vector. Combining the possible structures of Lorentz indices for electroweak gauge bosons, we obtain only one candidate
\begin{align}
U_{(S=1)}^{\alpha  \beta  \mu  \nu}  
&=  
\eta_\perp^{\alpha  \mu}  \eta_\perp^{\beta  \nu}  -  \eta_\perp^{\alpha  \nu}  \eta_\perp^{\beta  \mu}.
\end{align}
For $S  =  2$, we contract $\tilde{\Upsilon}_k  \tilde{\Upsilon}_\ell$, which corresponds to $\tilde{\Upsilon}_\alpha  \tilde{\Upsilon}_\beta$, with electroweak bosons. The possible structures are 
\begin{align}
U_{(S=2)}^{\alpha  \beta  \mu  \nu}  
&=  
\left\{
\begin{array}{l}
\eta_\perp^{\alpha  \mu}  \eta_\perp^{\beta  \nu}  +  \eta_\perp^{\alpha  \nu}  \eta_\perp^{\beta  \mu},
\\
(n_-  -  n_+)^\alpha  (n_-  -  n_+)^\beta  \eta_\perp^{\mu  \nu},
\\
\left[ 
(n_-  +  n_+)_\rho  (n_-  -  n_+)^{\alpha}  \eta_{u,  \sigma}^{\beta}
+
(n_-  +  n_+)_\rho  (n_-  -  n_+)^{\beta}  \eta_{u,  \sigma}^{\alpha}
\right]
\epsilon^{\rho  \sigma  \mu  \nu},
\\
\\
(n_-  +  n_+)_\rho  (n_-  -  n_+)_\sigma  \epsilon^{\alpha  \beta   \rho  \sigma}  \eta_\perp^{\mu  \nu},
\\
\left[ 
(n_-  +  n_+)_\rho  (n_-  -  n_+)^{\alpha}  \eta_{u,  \sigma}^{\beta}
-
(n_-  +  n_+)_\rho  (n_-  -  n_+)^{\beta}  \eta_{u,  \sigma}^{\alpha}
\right]
\epsilon^{\rho  \sigma  \mu  \nu}.
\end{array}
\right.
\label{eq:US2-npnm}
\end{align}
The first three (last two)  structures are symmetric (anti-symmetric) under $\alpha  \leftrightarrow  \beta$. All the other structures are power suppressed in the NR limit.

We can further restrict the structure of operators following the discussion in Ref.~\cite{Beneke:2019vhz}. In particular, the tensor structures for each operator should have certain properties by considering the symmetry expected in the EFT. First, we expect Bose symmetry in both the DM sector and the electroweak gauge field sector. For the DM sector, we can find the constraint by changing the label $A  \leftrightarrow  B$, $\alpha  \leftrightarrow  \beta$ and requiring the operator should be invariant. 
This brings us the following relation 
\begin{align}
U_{(S)}^{\alpha  \beta  \mu  \nu}  T_{\cal  J}^{ABCD}  =  U_{(S)}^{\beta  \alpha 
 \mu  \nu}  T_{\cal  J}^{BACD}.
\end{align}
For the electroweak gauge bosons, we can do the same analysis by changing the parameters $\mu  \leftrightarrow  \nu$ and $C  \leftrightarrow  D$, $c  \leftrightarrow  \bar{c}$, $n_-  \leftrightarrow  n_+$, and $s  \leftrightarrow  t$.
We obtain 
\begin{align}
U_{(S)}^{\alpha  \beta  \mu  \nu}  T_{\cal  J}^{ABCD}  =  \left. U_{(S)}^{\alpha  \beta  \nu  \mu}  T_{\cal  J}^{ABDC} \right|_{n_- 
 \leftrightarrow  n_+}.
\end{align}
After imposing Bose symmetry, we find the last two structures of Eq.~\eqref{eq:US2-npnm} are excluded. At this stage, the combination with $T_{\cal  J}$ for each $U_{(S)}$ is fixed; the symmetric (anti-symmetric) structure under $\alpha  \leftrightarrow  \beta$ can be nonzero only if combined with $T_{{\cal  J}=0,2}$ ($T_{{\cal  J}=1}$). Thus, we focus on the structure of $U_{(S)}$ in the following.

Second, we impose discrete symmetry such as CP symmetry on the operators. The CP symmetry is not conserved in the electroweak theory but holds for the DM annihilation into electroweak bosons at the NLL accuracy as we considered in this paper. Hence, we may further restrict the operator forms. In particular, the Lorentz tensor should satisfy the following condition. 
\begin{align}
U_{\alpha  \beta  \mu  \nu (S)}  =  U_{(S)}^{\alpha  \beta \mu  \nu}  (-1)^p,
\end{align}
where $(-1)^p$ comes from the parity transformation of $U_{(S)}$. In the end, we specify the following structures that are all consistent with the above structures. 
\begin{align}
U^{\alpha  \beta  \mu  \nu}_{(S=0)}  &=  \eta_u^{\alpha  \beta}  \eta_\perp^{\mu  \nu},
\\
U^{\alpha  \beta  \mu  \nu}_{(S=1)}  &=  \eta_\perp^{\alpha  \mu}  \eta_\perp^{\beta  \nu}  -  \eta_\perp^{\alpha  \nu}  \eta_\perp^{\beta  \mu},
\\
U^{\alpha  \beta  \mu  \nu}_{(S=2)}  &=  
\left\{
\begin{array}{l}
\eta_\perp^{\alpha  \mu}  \eta_\perp^{\beta  \nu}  +  \eta_\perp^{\alpha  \nu}  \eta_\perp^{\beta  \mu},
\\
(n_+  -  n_+)^\alpha  (n_+  -  n_+)^\beta  \eta_\perp^{\mu  \nu}.
\end{array}
\right.
\end{align}
This result includes all the necessary operator basis to match the full theory amplitudes. Notice that the the coefficient for $S=1$ operator vanishes at the non-relativistic leading order while the nonzero coefficient may be induced at higher accuracy.

\section{Derivation of the Wilson coefficients}
\label{sec:Wilson_coefficients-derivation}

\subsection{From amplitude to Wilson coefficients}

\begin{figure}[htb]
	\centering
\begin{minipage}{0.95\hsize}
\begin{minipage}{0.27\hsize}
	\includegraphics[width=1\hsize]{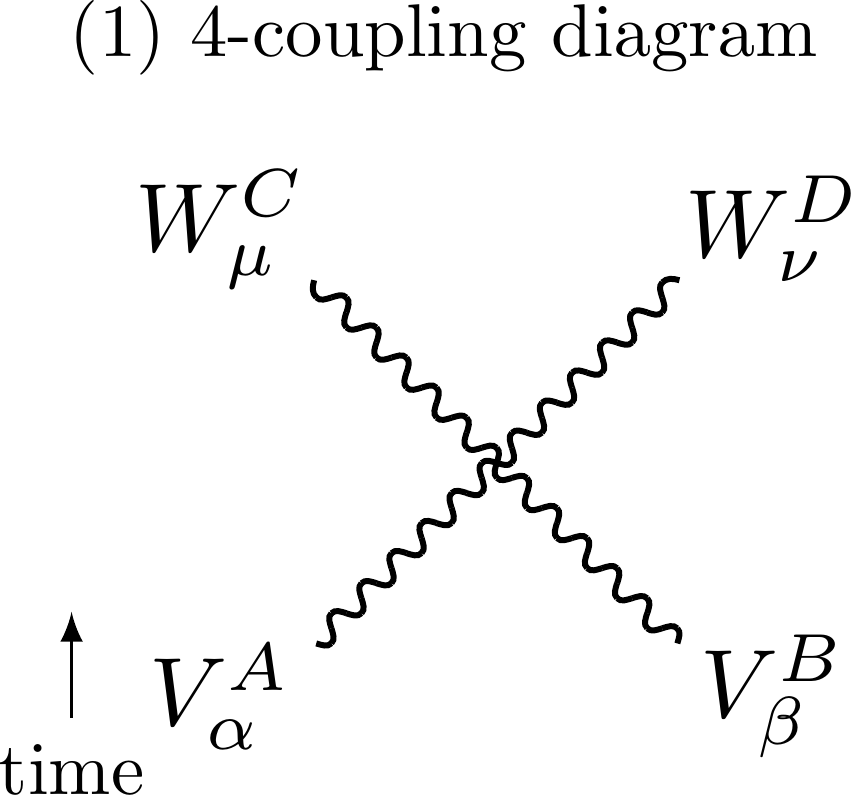}
\end{minipage}
	\qquad
\begin{minipage}{0.27\hsize}	
	\vspace{-0.5cm}
	\includegraphics[width=1\hsize]{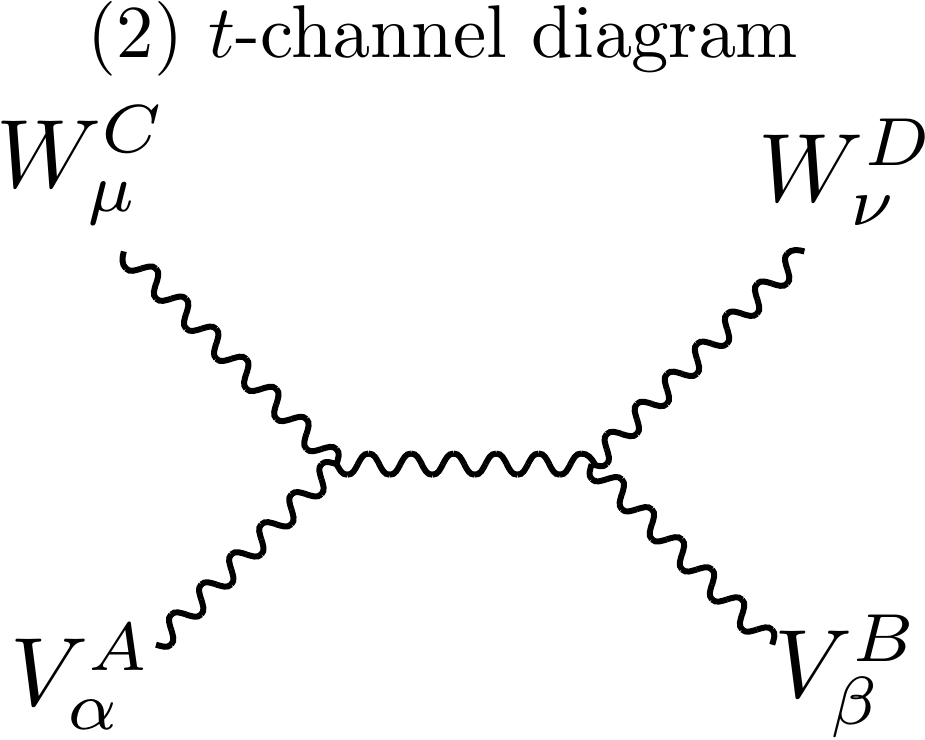}
\end{minipage}
	\qquad
\begin{minipage}{0.27\hsize}
	\vspace{-0.5cm}
	\includegraphics[width=1\hsize]{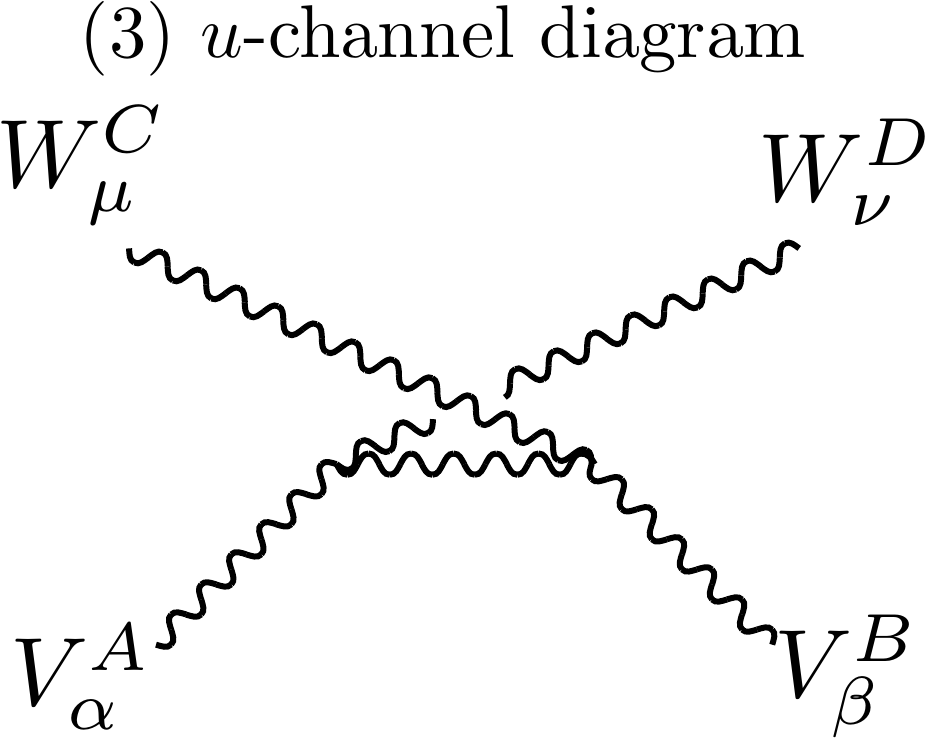}
\end{minipage}
\end{minipage}
	\caption{Diagrams for $V^A  V^B  \to  W^C  W^D$ in the unitary gauge. 
	}
	\label{fig:VVtoWWDiag4tu}
\end{figure}
%
We show the derivation of the Wilson coefficients in Eqs.~\eqref{eq:Wilsonini}-\eqref{eq:Wilsonfin}. First, we derive tree-level amplitudes of the $V^A  (p_1)  V^B  (p_2)  \to  W^C  (k_1)  W^D  (k_2)$ process in the full theory by taking the unitary gauge. The corresponding diagrams are shown in Fig.~\ref{fig:VVtoWWDiag4tu}. The relevant terms are the interaction term between $W$ and $V$ as read out from Eq.~\eqref{eq:LagSU(2)L}
\begin{align}
{\cal  L} 
&\supset  
-  \frac{i  g^2}{2}  
V_\alpha^A  V_\beta^B  W_\mu^C  W_\nu^D
\nn
\\
&\qquad
\times
\left[
\eta^{\alpha  \mu}  \eta^{\beta  \nu}
\left( \delta_{AC}  \delta_{BD}  -  \delta_{AD}  \delta_{BC} \right)
+
\left( 
\eta^{\alpha  \beta}  \eta^{\mu  \nu}
-  \eta^{\alpha  \nu}  \eta^{\beta  \mu}
\right)
\left( 
\delta_{AB}  \delta_{CD}
-  \delta_{AD}  \delta_{BC}
\right)
\right]
\nn
\\
&\qquad
-g  \epsilon^{abc}
\left[ 
\left( \del_\alpha  V_\beta^{\alpha  A} \right)  V^{\alpha  B}  W^{\beta  C} 
+  V_\alpha^A  \left( \del_\beta  V^{\alpha  B} \right)  W^{\beta  C}
+  V_\alpha^A  V_\beta^B  \left( \del^\alpha  W^{\beta  C} \right)
\right].
\end{align}
The amplitudes of quartic-coupling, $t$-channel, and $u$-channel diagrams are shown below\footnote{The $s$-channel diagrams do not induce the non-relativistic leading order effects. }
\begin{align}
i  {\cal  M}_4
&=
  i  g^2  \epsilon_\alpha^A  (p_1)  \epsilon_\beta^B  (p_2)  \epsilon^{*C}_\mu  (k_1)  \epsilon^{*D}_\nu  (k_2)
  \nn
  \\
  &\qquad
  \times  
  \Bigl[
        \delta_{AB}  \delta_{CD}  \left( \eta^{\alpha  \mu}  \eta^{\beta  \nu}  
  +  \eta^{\alpha  \nu}  \eta^{\beta  \mu}  -  2  \eta^{\mu  \nu}  \eta^{\alpha  \beta}   \right)   
    \nn
    \\
    &
    \qquad  \qquad
    +  \delta_{AC}  \delta_{BD}  \left( -  2  \eta^{\alpha  \mu}  \eta^{\beta  \nu}  +  \eta^{\alpha  \nu}  \eta^{\beta  \mu}  +  \eta^{\alpha  \beta}  \eta^{\mu  \nu} \right)
    \nn
    \\
    &
    \qquad  \qquad
    +  \delta_{AD}  \delta_{BC}  \left( \eta^{\alpha  \mu}  \eta^{\beta  \nu}    -  2    \eta^{\alpha  \nu}  \eta^{\beta  \mu}  +  \eta^{\alpha  \beta}  \eta^{\mu  \nu} \right)
  \Bigr],
\\
i  {\cal  M}_t
&=
g^2  
\epsilon_\alpha^A  (p_1)  \epsilon_\beta^B  (p_2)  \epsilon^{*C}_\mu  (k_1)  \epsilon^{*D}_\nu  (k_2)
\left( \delta^{AB}  \delta^{CD}  -  \delta^{AD}  \delta^{BC} \right)
\frac{-i  \left( \eta^{\rho  \sigma}  -  \frac{q^\rho  q^\sigma}{m_V^2} \right)}{q^2  -  m_V^2}
\\
&\quad
\times
\left[
-  \eta^{\alpha  \mu}  \left( p_1  +  k_1 \right)^\rho
+  \eta^{\alpha  \rho}  \left( p_1  +  q \right)^\mu
+  \eta^{\mu  \rho}  \left( k_1  -  q \right)^\alpha
\right]
\\
&\quad
\times
\left[
-  \eta^{\beta  \nu}  \left( p_2  +  k_2 \right)^\sigma
+  \eta^{\beta  \sigma}  \left( p_2  -  q \right)^\nu
+  \eta^{\sigma  \nu}  \left( k_2  +  q \right)^\beta
\right],
\\
i  {\cal  M}_u
&=
g^2  
\epsilon_\alpha^A  (p_1)  \epsilon_\beta^B  (p_2)  \epsilon^{*C}_\mu  (k_1)  \epsilon^{*D}_\nu  (k_2)
\left( \delta^{AB}  \delta^{CD}  -  \delta^{AC}  \delta^{BD} \right)
\frac{-i  \left( \eta^{\rho  \sigma}  -  \frac{q'^\rho  q'^\sigma}{m_V^2} \right)}{q'^2  -  m_V^2}
\\
&\quad
\times
\left[
-  \eta^{\alpha  \nu}  \left( p_1  +  k_2 \right)^\rho
+  \eta^{\alpha  \rho}  \left( p_1  +  q' \right)^\nu
+  \eta^{\nu  \rho}  \left( k_2  -  q' \right)^\alpha
\right]
\\
&\quad
\times
\left[
-  \eta^{\beta  \mu}  \left( p_2  +  k_1 \right)^\sigma
+  \eta^{\beta  \sigma}  \left( p_2  -  q' \right)^\mu
+  \eta^{\sigma  \mu}  \left( k_1  +  q' \right)^\beta
\right],
\end{align}
where $q  =  p_1  -  k_1$ and $q'  =  p_1  -  k_2$.

We can further simplify the amplitude using kinematical properties. Here, we take $x$-$z$ plane along to the momenta for the final state particles to simplify the expression. Then, the kinematics of the CM frame are defined below
\begin{align}
  p_{1\mu}  &=  \left( m_V,  \bm{p} \right)  
  &
  &\xrightarrow[]{\rm  L.O.}
  \left( m_V,  \bm{0} \right)  
  =
  m_V  u_\mu,
  \\
  p_{2\mu}  &=  \left( m_V,  -  \bm{p} \right)  
  &
  &\xrightarrow[]{\rm  L.O.}
  \left( m_V,  \bm{0} \right)  
  =
  m_V  u_\mu,
  \\
  k_{1\mu}  &=  \left( m_V,  \bm{k} \right)  
  &
  &\xrightarrow[]{\rm  L.O.}
  \left( m_V,   0,  0,  m_V  \right)   
  =
  m_V  \left( u_\mu  +  w_\mu \right),
  \\
  k_{2\mu}  &=  \left( m_V,  -  \bm{k} \right)  
  &
  &\xrightarrow[]{\rm  L.O.}
  \left( m_V,   0,  0,  -m_V  \right)  
  =
  m_V  \left( u_\mu  -  w_\mu \right).
\end{align}
We also show the leading order expressions by neglecting ${\cal O} (\vec{p})$ corrections, which is expressed by the following velocity vectors
\begin{align}
  u_\mu  &\equiv  \left( 1,  0,  0,  0 \right)  
  =
  \frac{n_+  +  n_-}{2},
  \label{eq:v_def}
  \\
  w_\mu  
  &\equiv  
  \left( 0,  0,  0,  1 \right)
  =
  \frac{n_+  -  n_-}{2}.
  \label{eq:w_def}
\end{align}
The initial $V$-particles have three polarization as shown below 
\begin{align}
  \epsilon  ( \lambda  =  +1 )
  &=
  \begin{pmatrix}
    0
    \\
    \frac{-1}{\sqrt{2}}
    \\
    \frac{i}{\sqrt{2}}
    \\
    0
  \end{pmatrix},
  &
  \epsilon  ( \lambda  =  -1 )
  &=
  \begin{pmatrix}
    0
    \\
    \frac{+1}{\sqrt{2}}
    \\
    \frac{i}{\sqrt{2}}
    \\
    0
  \end{pmatrix},
  &
  \epsilon  ( \lambda  =  0 )
  &=
  \begin{pmatrix}
    0
    \\
    0
    \\
    0
    \\
    1
  \end{pmatrix},
  \label{eq:initial_epsilon}
\end{align}
where $\lambda$ expresses the polarization of initial particles. The three vectors for spatial coordinates are the eigenvalue of $J_z$ of initial particles. 
\begin{align}
  J_z  &=  
  \begin{pmatrix}
    0  &  -i  &  0
    \\
    i  &  0  &  0
    \\
    0  &  0  &  0
  \end{pmatrix}.
\end{align}
The final state particles, which are massless in the heavy DM mass limit, have only transverse polarizations, and we chose the following choices. 
\begin{align}
  \epsilon  ( \lambda  =  \pm  1,  k_1 )
  &=
  \frac{1}{\sqrt{2}}
  \begin{pmatrix}
    0
    \\
    \mp  1
    \\
    -  i
    \\
    0
  \end{pmatrix},
  &
  \epsilon  ( \lambda  =  \pm  1,  k_2 )
  &=
  \frac{1}{\sqrt{2}}
  \begin{pmatrix}
    0
    \\
    \pm  1
    \\
    -  i
    \\
    0
  \end{pmatrix},
  \label{eq:final_epsilon}
\end{align}
Notice that the velocity vector $u$ cannot appear in the final expression because it is orthogonal with respect to all the other Lorentz vectors to be contracted.

Using these leading order expressions and the Ward-Takahashi identities, we can reduce the expression for $t$-channel and $u$-channel amplitude. 
\begin{align}
  i  {\cal  M}_{t}  +  i  {\cal  M}_{u}
  &=  
  i  g^2  \epsilon_\alpha^A  (p_1)  \epsilon_\beta^B  (p_2)  \epsilon^{*C}_\mu  (k_1)  \epsilon^{*D}_\nu  (k_2)
  \nn
  \\
  &
  \qquad
  \times  \Bigl[
  \delta_{AB}  \delta_{CD}  
  \left( -  4  w^\alpha  w^\beta  \eta^{\mu  \nu}  +  3  \eta^{\alpha  \mu}  \eta^{\beta  \nu}  +  3  \eta^{\alpha  \nu}  \eta^{\beta  \mu} \right)
  \nn
  \\
  &
  \qquad  \qquad
  +  \delta_{AC}  \delta_{BD}  \left( 2  w^\alpha  w^\beta  \eta^{\mu  \nu}  -  3  \eta^{\alpha  \nu}  \eta^{\beta  \mu} \right)
  \nn
  \\
  &
  \qquad  \qquad
  +  \delta_{AD}  \delta_{BD}  \left( 2  w^\alpha   w^\beta  \eta^{\mu  \nu}  -  3  \eta^{\alpha  \mu}  \eta^{\beta  \nu} \right)
  \Bigr].
\end{align}
Adding the total amplitude, we obtain
\begin{align}
  i  {\cal  M}_{\rm  tot}
  &=
  i  4  \pi  \alpha_2
  \epsilon_\rho^A  (p_1)  \epsilon_\sigma^B  (p_2)  \epsilon^{*C}_\mu  (k_1)  \epsilon^{*D}_\nu  (k_2)
  \left( \delta_{AC}  \delta_{BD}  +  \delta_{AD}  \delta_{BC}  -  2  \delta_{AB}  \delta_{CD} \right)
  \nn
  \\
  &
  \qquad
  \times
  \left[
  2  w^\alpha  w^\beta  \eta^{\mu  \nu}
  +  \eta^{\alpha  \beta}  \eta^{\mu  \nu}
  -2  \left( \eta^{\alpha  \mu}  \eta^{\beta  \nu}  +  \eta^{\alpha  \nu}  \eta^{\beta  \mu} \right)
  \right].
  \label{eq:Mtot}
\end{align}
Notice that the expression is already factorized for SU(2)$_L$ indices and the Lorentz indices, and symmetric under the exchange of the initial $V$-particles ($\alpha  \leftrightarrow  \beta$ and $A  \leftrightarrow  B$).

We match this amplitude in the full theory with the effective operators. It is convenient to use the irreducible operators, combining operator basis specified in Appendix.~\ref{sec:operators}. Since the amplitude in Eq.~\eqref{eq:Mtot} is symmetric under exchange of initial particles, we only need ${\cal  J}  =  0, 2$ and $S  =  0,  2$ for the matching. The relevant tensors to match the tree-level result, thus, are listed below:
\begin{align}
  &
  {\cal  J}  =  0
  &
  &
  :\quad
  T^{ABCD}_{{\cal  J}  =  0}  \equiv  \delta_{A  B}  \delta_{C  D},
  \\
  &
  {\cal  J}  =  2
  &
  &
  :\quad
  T^{ABCD}_{{\cal  J}  =  2}  \equiv  \delta_{AC}  \delta_{BD}  +  \delta_{AD}  \delta_{BC}  -  \frac{2}{3}  \delta_{AB}  \delta_{CD},
  \\
  \nn
  \\
  &
  S  =  0
  &
  &
  :\quad
  U^{\alpha  \beta  \mu  \nu}_{{\cal}  (S  =  0)}  \equiv  \eta^{\alpha  \beta}  \eta^{\mu  \nu},
  \\
  &
  S  =  2,~i=1
  &
  &
  :\quad
  U^{\alpha  \beta  \mu  \nu}_{{\cal}  (S  =  2)_1}  \equiv  \eta^{\alpha  \mu}   \eta^{\beta  \nu}  +  \eta^{\beta  \mu}  \eta^{\alpha  \nu}  -  \frac{2}{3}  \eta^{\alpha   \beta}  \eta^{\mu  \nu},
  \\
  &
  S  =  2,~i=2
  &
  &
  :\quad
  U^{\alpha  \beta  \mu  \nu}_{{\cal}  (S  =  2)_2}  \equiv  \left( n_+  -  n_- \right)^\alpha  \left( n_+  -  n_- \right)^\beta  \eta^{\mu  \nu}  +  \frac{4}{3}  \eta^{\alpha  \beta}  \eta^{\mu  \nu},
\end{align}
where we define ${\cal  J}, S  =  2$ tensor by subtracting ${\cal  J},  S  =  0$ tensor to form trace-less combination over the initial $V$-particles' indices. For $S  =  2$, we have two expressions for the structure of indices to match the amplitudes and we introduce label $i=1,2$ to distinguish them. Combining these tensors representing indices' structure, we obtain the following effective Lagrangian
\begin{align}
{\cal  L}_{\rm eff}
&=
\sum_{{\cal  J}  =  0,  2}  
{\int\diff t\diff s}\left( 
  C^{(0)}_{{\cal  J}}  {\cal  O}_{\cal  J}^{(S  =  0)}  
  +  C^{(2)_1}_{{\cal  J}}  {\cal  O}_{\cal  J}^{(S=2)_1}
  +  C^{(2)_2}_{{\cal  J}}  {\cal  O}_{\cal  J}^{(S=2)_2}
\right),
\end{align}
where
\begin{align}
{\cal  O}_{\cal  J}^{(S  =  0)}  
&=  
\tilde\vnr_\alpha^A  \tilde\vnr_\beta^B  \tilde\scA_{c\,\mu}^C  \tilde\scA_{\bar c\,\nu}^D
T^{ABCD}_{{\cal  J}} 
U^{\alpha  \beta  \mu  \nu}_{(S  =  0)},
\\
{\cal  O}_{\cal  J}^{(S  =  2)_1}  
&=  
\tilde\vnr_\alpha^A  \tilde\vnr_\beta^B  \tilde\scA_{c\,\mu}^C  \tilde\scA_{\bar c\,\nu}^D
T^{ABCD}_{{\cal  J}}
U^{\alpha  \beta  \mu  \nu}_{(S  =  2)_1},
\\
{\cal  O}_{\cal  J}^{(S  =  2)_2}  
&=
\tilde\vnr_\alpha^A  \tilde\vnr_\beta^B  \tilde\scA_{c\,\mu}^C  \tilde\scA_{\bar c\,\nu}^D
T^{ABCD}_{{\cal  J}}
U^{\alpha  \beta  \mu  \nu}_{(S  =  2)_2}.
\end{align}
These operators correspond to Eqs. \eqref{eq:opbasisi}-\eqref{eq:opbasisf} if particle fields and metrics are replaced into soft collinear effective fields and projected metric in Eqs.~\eqref{eq:metric_ini}-\eqref{eq:metric_fin}, respectively. The amplitude derived from the effective Lagrangian is 
\begin{align}
\label{eq:scetampl}
i  {\cal  M}_{\rm  eff}  
&=  i  2  C^{(S)_i}_{\cal  J}  
\epsilon_\alpha^A  (p_1)  \epsilon_\beta^B  (p_2)  \epsilon^{*C}_\mu  (k_1)  \epsilon^{*D}_\nu  (k_2)
T^{ABCD}_{{\cal  J}}
U^{\alpha  \beta  \mu  \nu}_{{(S)_i}},
\end{align}
where the factor of $2$ comes from the fact that the operator is invariant under the following label-change
\begin{align}
T^{ABCD}_{{\cal  J}}  
&=  T^{BACD}_{{\cal  J}},
&
U^{\alpha  \beta  \mu  \nu}_{(S)}  
&=  U^{\beta  \alpha  \mu  \nu}_{(S)}.
\end{align}
The Wilson coefficients are determined to realize ${\cal M}_{\rm  tot}  =  {\cal  M}_{\rm  eff}$, and the result is summarized in Eqs.~\eqref{eq:Wilsonini}-\eqref{eq:Wilsonfin}.

\subsection{Cross section at tree-level}

Expanding Eq.~\eqref{eq:factorizationII} at tree-level yields the following expression in matrix notation
\begin{equation}
    \frac{(\diff\tilde\sigma v)^{S=0}}{\diff E}=
    \begin{pmatrix}
        0 & 0 \\
        0 & 1
    \end{pmatrix}\frac{4\pi\alpha_2^2s_W^2}{3m_V^2}\delta(E-m_V)\quad ,\quad
    \frac{(\diff\tilde\sigma v)^{S=2}}{\diff E}=
    \begin{pmatrix}
        0 & 0 \\
        0 & 1
    \end{pmatrix}\frac{64\pi\alpha_2^2s_W^2}{9m_V^2}\delta(E-m_V)\ ,    
\end{equation}
which when integrated and added together reduces to the known result \cite{Abe:2021mry}
\begin{equation}
    (\tilde\sigma v)_{\gamma\gamma+\gamma Z}^{S\textrm{ even}}=
    \begin{pmatrix}
        0 & 0 \\
        0 & 1
    \end{pmatrix}\frac{2\times38\pi\alpha_2^2s_W^2}{9m_V^2}\ .    
\end{equation}

For the numerical computation of the continuum part of the spectrum shown in Fig.~\ref{fig:diffspec} we used the following annihilation matrices 
\begin{equation}
    (\tilde\sigma v)^{S=0}_{W^+W^-}=
    \begin{pmatrix}
        2 & \sqrt2 \\
        \sqrt2 & 1
    \end{pmatrix}\frac{\pi\alpha_2^2}{3m_V^2}\quad ,\quad
    (\tilde\sigma v)^{S=2}_{W^+W^-}=
    \begin{pmatrix}
        2 & \sqrt2 \\
        \sqrt2 & 1
    \end{pmatrix}\frac{16\pi\alpha_2^2}{9m_V^2}\ , 
\end{equation}
which adds up to
\begin{equation}
    (\tilde\sigma v)_{W^+W^-}^{S\textrm{ even}}=
    \begin{pmatrix}
        2 & \sqrt2 \\
        \sqrt2 & 1
    \end{pmatrix}\frac{19\pi\alpha_2^2}{9m_V^2}\ .
\end{equation}
We also note also the total cross section as 
\begin{equation}
    (\tilde\sigma v)_{\rm TOT}^{S\textrm{ even}}=
    \begin{pmatrix}
        2 & \sqrt2 \\
        \sqrt2 & 3
    \end{pmatrix}\frac{19\pi\alpha_2^2}{9m_V^2}\ . 
\end{equation}

\bibliographystyle{utphysmod}
\bibliography{references}

\providecommand{\href}[2]{#2}\begingroup\raggedright\begin{thebibliography}{10}

\bibitem{Planck:2018vyg}
{\bfseries Planck} Collaboration, {\em {Planck 2018 results. VI. Cosmological parameters}}, \href{https://dx.doi.org/10.1051/0004-6361/201833910}{Astron.\  Astrophys.\  {\bfseries 641} (2020) A6} {\ttfamily [\href{https://arxiv.org/abs/1807.06209}{arXiv:1807.06209}]}. [Erratum: Astron.Astrophys. 652, C4 (2021)].

\bibitem{Cirelli:2005uq}
M.~Cirelli, N.~Fornengo, and A.~Strumia, {\em {Minimal dark matter}}, \href{https://dx.doi.org/10.1016/j.nuclphysb.2006.07.012}{Nucl.\  Phys.\  B {\bfseries 753} (2006) 178--194} {\ttfamily [\href{https://arxiv.org/abs/hep-ph/0512090}{hep-ph/0512090}]}.

\bibitem{Cirelli:2007xd}
M.~Cirelli, A.~Strumia, and M.~Tamburini, {\em {Cosmology and Astrophysics of Minimal Dark Matter}}, \href{https://dx.doi.org/10.1016/j.nuclphysb.2007.07.023}{Nucl.\  Phys.\  B {\bfseries 787} (2007) 152--175} {\ttfamily [\href{https://arxiv.org/abs/0706.4071}{arXiv:0706.4071}]}.

\bibitem{Cirelli:2009uv}
M.~Cirelli and A.~Strumia, {\em {Minimal Dark Matter: Model and results}}, \href{https://dx.doi.org/10.1088/1367-2630/11/10/105005}{New J.\  Phys.\  {\bfseries 11} (2009) 105005} {\ttfamily [\href{https://arxiv.org/abs/0903.3381}{arXiv:0903.3381}]}.

\bibitem{Flacke:2008ne}
T.~Flacke, A.~Menon, and D.~J.~Phalen, {\em {Non-minimal universal extra dimensions}}, \href{https://dx.doi.org/10.1103/PhysRevD.79.056009}{Phys.\  Rev.\  D {\bfseries 79} (2009) 056009} {\ttfamily [\href{https://arxiv.org/abs/0811.1598}{arXiv:0811.1598}]}.

\bibitem{Flacke:2017xsv}
T.~Flacke, D.~W.~Kang, K.~Kong, G.~Mohlabeng, and S.~C.~Park, {\em {Electroweak Kaluza-Klein Dark Matter}}, \href{https://dx.doi.org/10.1007/JHEP04(2017)041}{JHEP {\bfseries 04} (2017) 041} {\ttfamily [\href{https://arxiv.org/abs/1702.02949}{arXiv:1702.02949}]}.

\bibitem{Maru:2018ocf}
N.~Maru, N.~Okada, and S.~Okada, {\em {$SU(2)_L$ doublet vector dark matter from gauge-Higgs unification}}, \href{https://dx.doi.org/10.1103/PhysRevD.98.075021}{Phys.\  Rev.\  D {\bfseries 98} (2018) 075021} {\ttfamily [\href{https://arxiv.org/abs/1803.01274}{arXiv:1803.01274}]}.

\bibitem{Arkani-Hamed:2001kyx}
N.~Arkani-Hamed, A.~G.~Cohen, and H.~Georgi, {\em {(De)constructing dimensions}}, \href{https://dx.doi.org/10.1103/PhysRevLett.86.4757}{Phys.\  Rev.\  Lett.\  {\bfseries 86} (2001) 4757--4761} {\ttfamily [\href{https://arxiv.org/abs/hep-th/0104005}{hep-th/0104005}]}.

\bibitem{Arkani-Hamed:2001nha}
N.~Arkani-Hamed, A.~G.~Cohen, and H.~Georgi, {\em {Electroweak symmetry breaking from dimensional deconstruction}}, \href{https://dx.doi.org/10.1016/S0370-2693(01)00741-9}{Phys.\  Lett.\  B {\bfseries 513} (2001) 232--240} {\ttfamily [\href{https://arxiv.org/abs/hep-ph/0105239}{hep-ph/0105239}]}.

\bibitem{Abe:2020mph}
T.~Abe, M.~Fujiwara, J.~Hisano, and K.~Matsushita, {\em {A model of electroweakly interacting non-abelian vector dark matter}}, \href{https://dx.doi.org/10.1007/JHEP07(2020)136}{JHEP {\bfseries 07} (2020) 136} {\ttfamily [\href{https://arxiv.org/abs/2004.00884}{arXiv:2004.00884}]}.

\bibitem{Abe:2021mry}
T.~Abe, M.~Fujiwara, J.~Hisano, and K.~Matsushita, {\em {Gamma-ray line from electroweakly interacting non-abelian spin-1 dark matter}}, \href{https://dx.doi.org/10.1007/JHEP10(2021)163}{JHEP {\bfseries 10} (2021) 163} {\ttfamily [\href{https://arxiv.org/abs/2107.10029}{arXiv:2107.10029}]}.

\bibitem{Hryczuk:2011vi}
A.~Hryczuk and R.~Iengo, {\em {The one-loop and Sommerfeld electroweak corrections to the Wino dark matter annihilation}}, \href{https://dx.doi.org/10.1007/JHEP01(2012)163}{JHEP {\bfseries 01} (2012) 163} {\ttfamily [\href{https://arxiv.org/abs/1111.2916}{arXiv:1111.2916}]}. [Erratum: JHEP 06, 137 (2012)].

\bibitem{Cohen:2013ama}
T.~Cohen, M.~Lisanti, A.~Pierce, and T.~R.~Slatyer, {\em {Wino Dark Matter Under Siege}}, \href{https://dx.doi.org/10.1088/1475-7516/2013/10/061}{JCAP {\bfseries 10} (2013) 061} {\ttfamily [\href{https://arxiv.org/abs/1307.4082}{arXiv:1307.4082}]}.

\bibitem{Baumgart:2014vma}
M.~Baumgart, I.~Z.~Rothstein, and V.~Vaidya, {\em {Calculating the Annihilation Rate of Weakly Interacting Massive Particles}}, \href{https://dx.doi.org/10.1103/PhysRevLett.114.211301}{Phys.\  Rev.\  Lett.\  {\bfseries 114} (2015) 211301} {\ttfamily [\href{https://arxiv.org/abs/1409.4415}{arXiv:1409.4415}]}.

\bibitem{Bauer:2014ula}
M.~Bauer, T.~Cohen, R.~J.~Hill, and M.~P.~Solon, {\em {Soft Collinear Effective Theory for Heavy WIMP Annihilation}}, \href{https://dx.doi.org/10.1007/JHEP01(2015)099}{JHEP {\bfseries 01} (2015) 099} {\ttfamily [\href{https://arxiv.org/abs/1409.7392}{arXiv:1409.7392}]}.

\bibitem{Baumgart:2018yed}
M.~Baumgart, {\em et al.}, {\em {Precision Photon Spectra for Wino Annihilation}}, \href{https://dx.doi.org/10.1007/JHEP01(2019)036}{JHEP {\bfseries 01} (2019) 036} {\ttfamily [\href{https://arxiv.org/abs/1808.08956}{arXiv:1808.08956}]}.

\bibitem{Beneke:2019vhz}
M.~Beneke, A.~Broggio, C.~Hasner, K.~Urban, and M.~Vollmann, {\em {Resummed photon spectrum from dark matter annihilation for intermediate and narrow energy resolution}}, \href{https://dx.doi.org/10.1007/JHEP08(2019)103}{JHEP {\bfseries 08} (2019) 103} {\ttfamily [\href{https://arxiv.org/abs/1903.08702}{arXiv:1903.08702}]}. [Erratum: JHEP 07, 145 (2020)].

\bibitem{Beneke:2020vff}
M.~Beneke, R.~Szafron, and K.~Urban, {\em {Sommerfeld-corrected relic abundance of wino dark matter with NLO electroweak potentials}}, \href{https://dx.doi.org/10.1007/JHEP02(2021)020}{JHEP {\bfseries 02} (2021) 020} {\ttfamily [\href{https://arxiv.org/abs/2009.00640}{arXiv:2009.00640}]}.

\bibitem{Pappadopulo:2014qza}
D.~Pappadopulo, A.~Thamm, R.~Torre, and A.~Wulzer, {\em {Heavy Vector Triplets: Bridging Theory and Data}}, \href{https://dx.doi.org/10.1007/JHEP09(2014)060}{JHEP {\bfseries 09} (2014) 060} {\ttfamily [\href{https://arxiv.org/abs/1402.4431}{arXiv:1402.4431}]}.

\bibitem{Hisano:2004ds}
J.~Hisano, S.~Matsumoto, M.~M.~Nojiri, and O.~Saito, {\em {Non-perturbative effect on dark matter annihilation and gamma ray signature from galactic center}}, \href{https://dx.doi.org/10.1103/PhysRevD.71.063528}{Phys.\  Rev.\  D {\bfseries 71} (2005) 063528} {\ttfamily [\href{https://arxiv.org/abs/hep-ph/0412403}{hep-ph/0412403}]}.

\bibitem{Arkani-Hamed:2008hhe}
N.~Arkani-Hamed, D.~P.~Finkbeiner, T.~R.~Slatyer, and N.~Weiner, {\em {A Theory of Dark Matter}}, \href{https://dx.doi.org/10.1103/PhysRevD.79.015014}{Phys.\  Rev.\  D {\bfseries 79} (2009) 015014} {\ttfamily [\href{https://arxiv.org/abs/0810.0713}{arXiv:0810.0713}]}.

\bibitem{Ovanesyan:2014fwa}
G.~Ovanesyan, T.~R.~Slatyer, and I.~W.~Stewart, {\em {Heavy Dark Matter Annihilation from Effective Field Theory}}, \href{https://dx.doi.org/10.1103/PhysRevLett.114.211302}{Phys.\  Rev.\  Lett.\  {\bfseries 114} (2015) 211302} {\ttfamily [\href{https://arxiv.org/abs/1409.8294}{arXiv:1409.8294}]}.

\bibitem{Beneke:2018ssm}
M.~Beneke, A.~Broggio, C.~Hasner, and M.~Vollmann, {\em {Energetic $\gamma$-rays from TeV scale dark matter annihilation resummed}}, \href{https://dx.doi.org/10.1016/j.physletb.2018.10.008}{Phys.\  Lett.\  B {\bfseries 786} (2018) 347--354} {\ttfamily [\href{https://arxiv.org/abs/1805.07367}{arXiv:1805.07367}]}. [Erratum: Phys.Lett.B 810, 135831 (2020)].

\bibitem{Ovanesyan:2016vkk}
G.~Ovanesyan, N.~L.~Rodd, T.~R.~Slatyer, and I.~W.~Stewart, {\em {One-loop correction to heavy dark matter annihilation}}, \href{https://dx.doi.org/10.1103/PhysRevD.95.055001}{Phys.\  Rev.\  D {\bfseries 95} (2017) 055001} {\ttfamily [\href{https://arxiv.org/abs/1612.04814}{arXiv:1612.04814}]}. [Erratum: Phys.Rev.D 100, 119901 (2019)].

\bibitem{Baumgart:2017nsr}
M.~Baumgart, {\em et al.}, {\em {Resummed Photon Spectra for WIMP Annihilation}}, \href{https://dx.doi.org/10.1007/JHEP03(2018)117}{JHEP {\bfseries 03} (2018) 117} {\ttfamily [\href{https://arxiv.org/abs/1712.07656}{arXiv:1712.07656}]}.

\bibitem{Beneke:2019gtg}
M.~Beneke, C.~Hasner, K.~Urban, and M.~Vollmann, {\em {Precise yield of high-energy photons from Higgsino dark matter annihilation}}, \href{https://dx.doi.org/10.1007/JHEP03(2020)030}{JHEP {\bfseries 03} (2020) 030} {\ttfamily [\href{https://arxiv.org/abs/1912.02034}{arXiv:1912.02034}]}.

\bibitem{Beneke:2014gja}
M.~Beneke, C.~Hellmann, and P.~Ruiz-Femenia, {\em {Non-relativistic pair annihilation of nearly mass degenerate neutralinos and charginos III. Computation of the Sommerfeld enhancements}}, \href{https://dx.doi.org/10.1007/JHEP05(2015)115}{JHEP {\bfseries 05} (2015) 115} {\ttfamily [\href{https://arxiv.org/abs/1411.6924}{arXiv:1411.6924}]}.

\bibitem{Landau:1948kw}
L.~D.~Landau, {\em {On the angular momentum of a system of two photons}}, \href{https://dx.doi.org/10.1016/B978-0-08-010586-4.50070-5}{Dokl.\  Akad.\  Nauk SSSR {\bfseries 60} (1948) 207--209}.

\bibitem{Yang:1950rg}
C.-N.~Yang, {\em {Selection Rules for the Dematerialization of a Particle Into Two Photons}}, \href{https://dx.doi.org/10.1103/PhysRev.77.242}{Phys.\  Rev.\  {\bfseries 77} (1950) 242--245}.

\bibitem{Yamada:2009ve}
Y.~Yamada, {\em {Electroweak two-loop contribution to the mass splitting within a new heavy SU(2)(L) fermion multiplet}}, \href{https://dx.doi.org/10.1016/j.physletb.2009.11.044}{Phys.\  Lett.\  B {\bfseries 682} (2010) 435--440} {\ttfamily [\href{https://arxiv.org/abs/0906.5207}{arXiv:0906.5207}]}.

\bibitem{Ibe:2012sx}
M.~Ibe, S.~Matsumoto, and R.~Sato, {\em {Mass Splitting between Charged and Neutral Winos at Two-Loop Level}}, \href{https://dx.doi.org/10.1016/j.physletb.2013.03.015}{Phys.\  Lett.\  B {\bfseries 721} (2013) 252--260} {\ttfamily [\href{https://arxiv.org/abs/1212.5989}{arXiv:1212.5989}]}.

\bibitem{McKay:2017xlc}
J.~McKay and P.~Scott, {\em {Two-loop mass splittings in electroweak multiplets: winos and minimal dark matter}}, \href{https://dx.doi.org/10.1103/PhysRevD.97.055049}{Phys.\  Rev.\  D {\bfseries 97} (2018) 055049} {\ttfamily [\href{https://arxiv.org/abs/1712.00968}{arXiv:1712.00968}]}.

\bibitem{Beneke:2019qaa}
M.~Beneke, R.~Szafron, and K.~Urban, {\em {Wino potential and Sommerfeld effect at NLO}}, \href{https://dx.doi.org/10.1016/j.physletb.2019.135112}{Phys.\  Lett.\  B {\bfseries 800} (2020) 135112} {\ttfamily [\href{https://arxiv.org/abs/1909.04584}{arXiv:1909.04584}]}.

\bibitem{Beneke:2022eci}
M.~Beneke, K.~Urban, and M.~Vollmann, {\em {Matching resummed endpoint and continuum \ensuremath{\gamma}-ray spectra from dark-matter annihilation}}, \href{https://dx.doi.org/10.1016/j.physletb.2022.137248}{Phys.\  Lett.\  B {\bfseries 834} (2022) 137248} {\ttfamily [\href{https://arxiv.org/abs/2203.01692}{arXiv:2203.01692}]}.

\bibitem{Fischer:2016vfv}
N.~Fischer, S.~Prestel, M.~Ritzmann, and P.~Skands, {\em {Vincia for Hadron Colliders}}, \href{https://dx.doi.org/10.1140/epjc/s10052-016-4429-6}{Eur.\  Phys.\  J.\  C {\bfseries 76} (2016) 589} {\ttfamily [\href{https://arxiv.org/abs/1605.06142}{arXiv:1605.06142}]}.

\bibitem{Arina:2023eic}
C.~Arina, {\em et al.}, {\em {CosmiXs: cosmic messenger spectra for indirect dark matter searches}}, \href{https://dx.doi.org/10.1088/1475-7516/2024/03/035}{JCAP {\bfseries 03} (2024) 035} {\ttfamily [\href{https://arxiv.org/abs/2312.01153}{arXiv:2312.01153}]}.

\end{thebibliography}\endgroup

\end{document}